\documentclass[prb,floatfix,superscriptaddress]{revtex4}
\usepackage{graphicx}
\usepackage{verbatim}
\usepackage{times}
\usepackage{subfigure}
\usepackage{amsmath}
\usepackage{natbib}

\begin{document}

\title{Pressure-energy correlations in liquids. III. Statistical mechanics and thermodynamics of liquids with hidden scale invariance}

\author{Thomas B. Schr{\o}der}
\email{tbs@ruc.dk}
\affiliation{DNRF Center ``Glass and Time'', IMFUFA, Dept. of Sciences, 
Roskilde University, P.O. Box 260, DK-4000 Roskilde, Denmark}
\author{Nicholas P. Bailey}
\email{nbailey@ruc.dk}
\affiliation{DNRF Center ``Glass and Time'', IMFUFA, Dept. of Sciences, 
Roskilde University, P.O. Box 260, DK-4000 Roskilde, Denmark}
\author{Ulf R. Pedersen}
\email{urp@ruc.dk}
\affiliation{DNRF Center ``Glass and Time'', IMFUFA, Dept. of Sciences, 
Roskilde University, P.O. Box 260, DK-4000 Roskilde, Denmark}
\author{Nicoletta Gnan}
\email{ngnan@ruc.dk}
\affiliation{DNRF Center ``Glass and Time'', IMFUFA, Dept. of Sciences, 
Roskilde University, P.O. Box 260, DK-4000 Roskilde, Denmark}
\author{ Jeppe C. Dyre }
\email{dyre@ruc.dk}
\affiliation{DNRF Center ``Glass and Time'', IMFUFA, Dept. of Sciences, 
Roskilde University, P.O. Box 260, DK-4000 Roskilde, Denmark}

\begin{abstract}
In this third paper of the series, which started with [N. P. Bailey {\it et al.}, J. Chem. Phys. {\bf 129}, 184507 and 184508 (2008)], we continue the development of the theoretical understanding of  strongly correlating liquids -- those whose instantaneous potential energy and virial are strongly correlated in their thermal equilibrium fluctuations at constant volume. The existence of such liquids was detailed in previous work which identified them, based on computer simulations, as a large class of liquids, including van der Waals liquids but not, e.g., hydrogen-bonded liquids. We here discuss the following: (1) The scaling properties of inverse power-law and extended inverse power-law potentials (the latter include a linear term which ``hides'' the approximate scale invariance); (2) results from computer simulations of molecular models concerning out-of-equilibrium conditions; (3) ensemble dependence of the virial / potential energy correlation coefficient; (4) connection to the Gr\"uneisen parameter; (5) interpretation of strong correlations in terms of the energy-bond formalism.
\end{abstract}

\newcommand{\angleb}[1]{\langle #1 \rangle}
\newcommand{\nod}{\noindent}
\newcommand{\half}{\frac{1}{2}}
\newcommand{\bfa}[1]{\mathbf{#1}} 

\newcommand{\la}{\langle}
\newcommand{\ra}{\rangle}

\date{\today}

\maketitle

\section{Introduction}

\begin{figure}
\includegraphics[width=10cm]{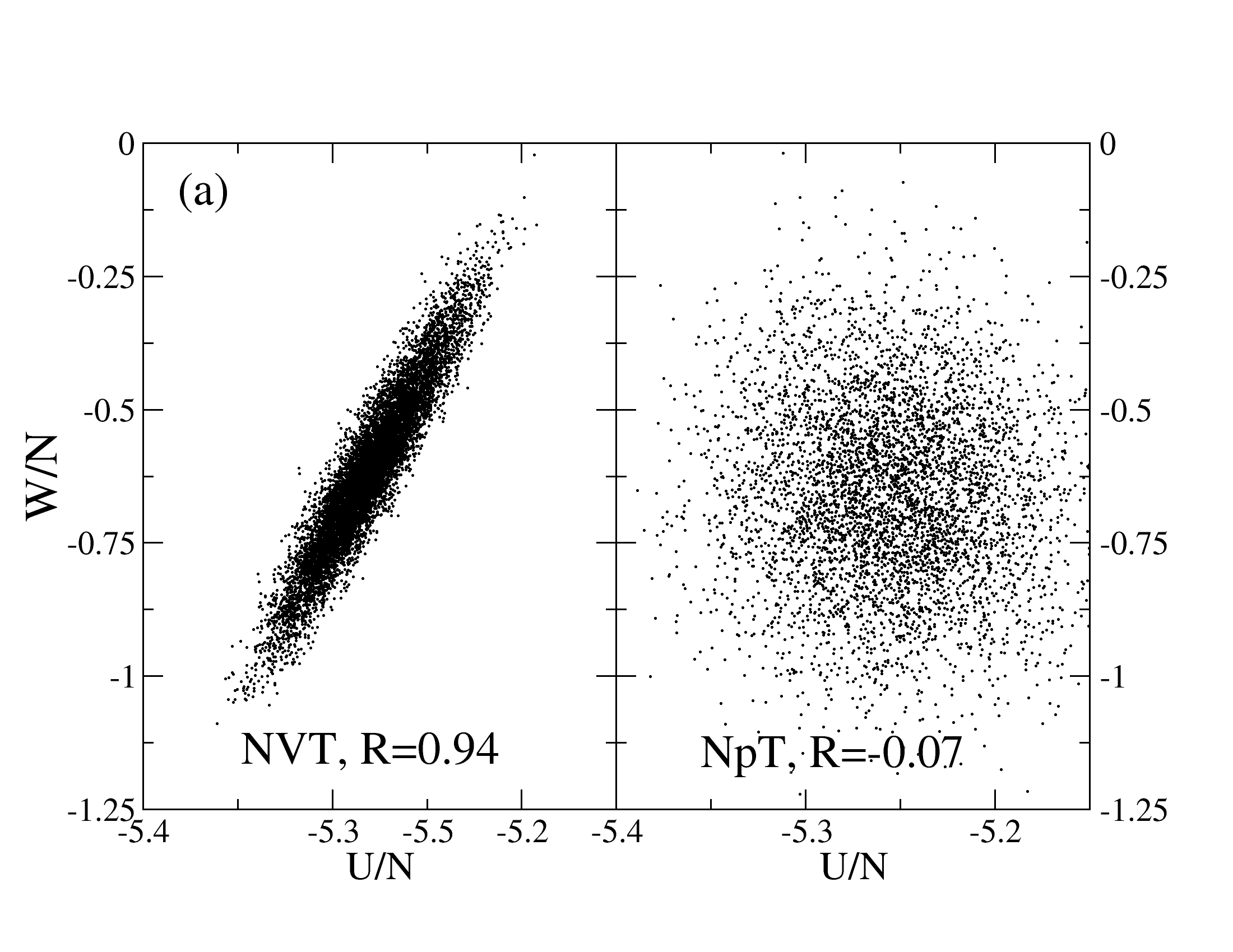}
\includegraphics[width=10cm]{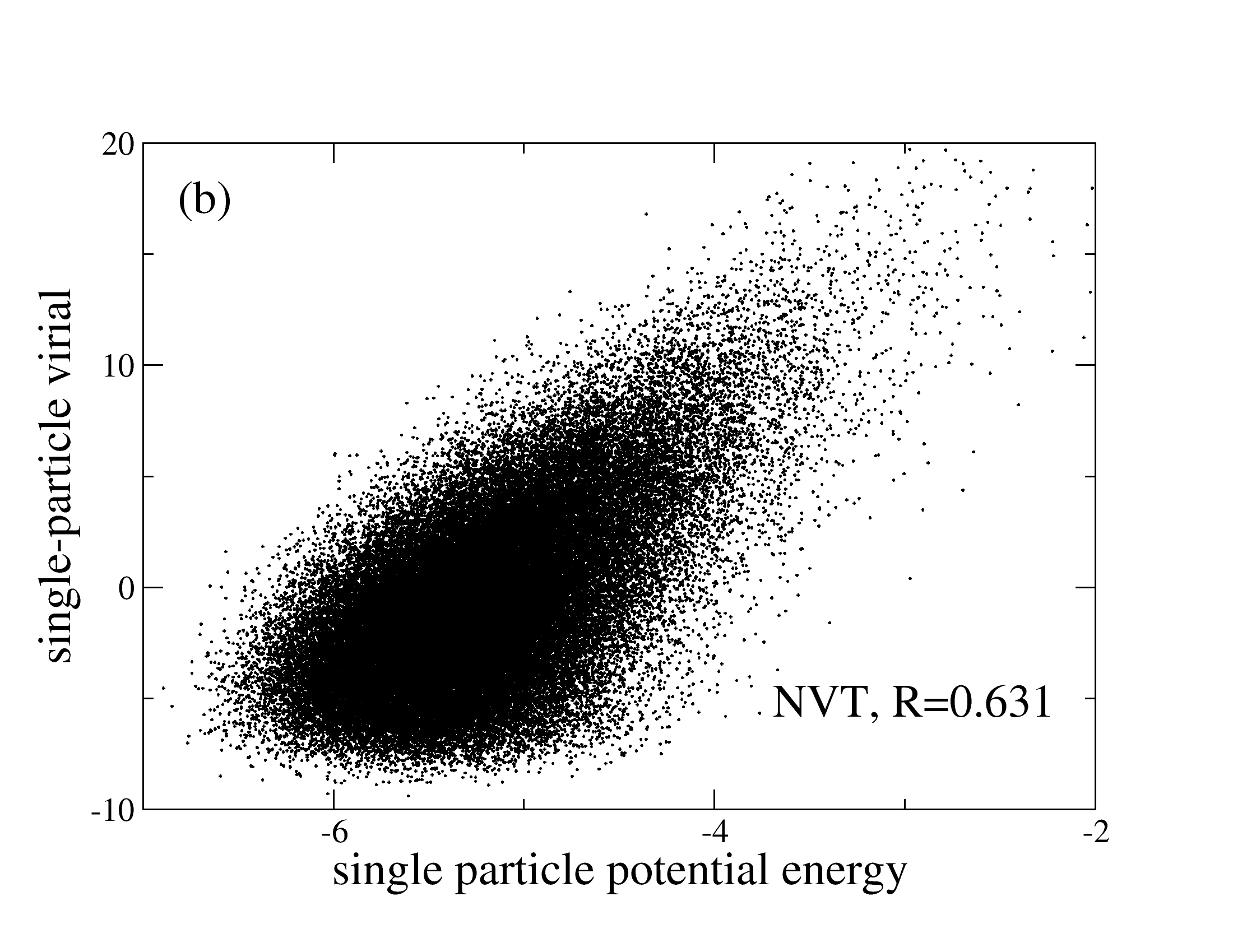}
\caption{\label{WU_NVT_NVP} (a) Scatter plot of total virial and potential energy (in Lennard-Jones units) for the standard single-component LJ liquid at $T=80$K (Argon units) and near-zero pressure,  simulated at constant volume (density $\rho=34.6$ mol/l, Argon units, left panel) and constant pressure (1.5 MPa, Argon units, right panel). 
(b)  Scatter plot of single-particle virial and potential energy for the same simulation as in the left panel of (a). The single-particle $WU$ correlation is much weaker, $R=0.63$, showing that collective effects are crucial for the correlation.}
\end{figure}

In a series of papers published last year \cite{Pedersen/others:2008,Pedersen/others:2008a,Bailey/others:2008a, Bailey/others:2008b,Bailey/others:2008c} we introduced the concept of strongly correlating liquids and demonstrated by computer simulations that this includes a large class of model liquids. Specifically, the fluctuations which are in many cases strongly correlated are those of the configurational parts of pressure and energy, i.e., the parts in addition to the ideal gas terms, coming from the interatomic forces. Recall that for any microscopic state, energy $E$ and pressure $p$ have contributions both from particle momenta and positions:

\begin{align}
E &= K(\bfa{p}_1,\ldots,\bfa{p}_N) + U(\bfa{r}_1,\ldots,\bfa{r}_N) \nonumber\\ 
 p &= Nk_BT(\bfa{p}_1,\ldots,\bfa{p}_N)/V+W(\bfa{r}_1,\ldots,\bfa{r}_N)/V\,.
\end{align}
Here $K$ and $U$ are the kinetic and potential energies, respectively, and $T(\bfa{p}_1,\ldots,\bfa{p}_N)$ is the ``kinetic temperature'', proportional to the kinetic energy per particle.\cite{Allen/Tildesley:1987}  The configurational  contribution to pressure is the virial $W$, which is defined\cite{Allen/Tildesley:1987} by

\begin{equation}\label{generalWformula}
W = -\frac{1}{3} \sum_i \bfa{r}_i \cdot \bfa{\nabla}_{\bfa{r}_i} U\,.
\end{equation}
For a liquid with pair interactions, if $v(r)$ is the pair potential and $r_{ij}$ is the distance between particles $i$ and $j$, we have

\begin{align}
U_{\textrm{pair}} &= \sum_{i<j} v(r_{ij}) \\
W_{\textrm{pair}} &= -\frac{1}{3}\sum_{i<j} r_{ij} v'(r_{ij})\,.
\end{align}
Strong $W,U$ correlation, if present at all, is observed under conditions of fixed volume, as illustrated in Fig.~\ref{WU_NVT_NVP}(a). The degree of correlation is quantified by the standard correlation coefficient $R$, defined\cite{Pedersen/others:2008, Bailey/others:2008b} by

\begin{equation}\label{correlationCoeff}
R=\frac {\langle\Delta W \Delta U\rangle}{\sqrt{\langle(\Delta W)^2\rangle}\sqrt{\langle(\Delta U)^2\rangle}}\,.
\end{equation}
Here and henceforth, unless otherwise specified, angle brackets $\langle\rangle$ denote thermal NVT ensemble averages; $\Delta$ denotes deviation from the average value of the quantity in question. We call liquids with $R>0.9$ strongly correlating. Another characteristic quantity is the ``slope'' $\gamma$, which we define\cite{Pedersen/others:2008,Bailey/others:2008b} as the ratio of standard deviations:

\begin{equation}\label{slopeDefinition}
\gamma = \frac{\sqrt{\angleb{(\Delta W)^2}}} {\sqrt{\angleb{(\Delta U)^2}}}\,.
\end{equation}
In the limit of perfect correlation ($R\rightarrow1$) $\gamma$ becomes equal to the standard linear-regression slope for $W$ as a function of $U$ at fixed volume. 

In Paper I\cite{Bailey/others:2008b} of this series it was shown that strongly correlating liquids are typically those with van der Waals type attraction and steep repulsion, which in simulations are often modelled by combinations of one or more Lennard-Jones type potentials. Typical slope values for the latter are of order 6, depending slightly on state point (in the limit of very high density or temperature the slope converges slowly to 4). Experimental data for Argon were analyzed and shown to be consistent with strong correlations ($R>0.95$) in the region of the phase diagram where quantum effects are not important.

It is worth noting that the class of strongly correlating liquids does not simply correspond to radially symmetric pair potentials. Firstly, two metallic systems with many-body potentials were found to be strongly correlating; it is probably true for metallic systems in general, although this needs to be confirmed. Also many molecular liquids are strongly correlating. In fact, any potential with an inverse power-law dependence on distances (not necessarily based on pair interactions) is perfectly correlating. Secondly, there exist radially symmetric pair potentials which are not strongly correlating, for example the Dzugutov system.\cite{Bailey/others:2008b,Dzugutov:1992} One reason for strong correlation not to hold in some molecular systems is the presence of Coulombic terms in the potential. By themselves these would give strong correlation, but their combination with Lennard-Jones forces typically leads to weak correlation. This was detailed in Paper I, which presented results from simulations of 13 different model liquids. In our present understanding based on these simulations, liquids with two length scales in their potentials are rarely strongly correlating.

Paper II\cite{Bailey/others:2008c} in this series analyzed the case of the standard single-component Lennard-Jones liquid in detail. Building on the fact that inverse power-law potentials $\propto r^{-n}$ are perfectly correlating, the results of this analysis can be summarized as follows: (1) Almost all of the fluctuations in $W$ and $U$ come from interparticle separations in the region of the first peak of the radial distribution function $g(r)$; (2) in this region the Lennard-Jones potential is approximated very well by the sum of an inverse power law with exponent $n\sim$18 and a linear term $B+Cr$; (3) when volume is fixed, the parts of $W$ and $U$ that come from the linear term are almost constant. Our initial and simpler explanation of strong $WU$ correlations (Ref.~\onlinecite{Pedersen/others:2008}) was based on the dominance of close encounters, i.e., that it is only the nature of the repulsive part of the potential that matters for the strong correlations. This explanation, however, is adequate at high pressure / density only. It does not explain the requirement of fixed volume, nor the  fact that strong correlation is observed even at zero pressure, as well as for the low-temperature / low-pressure (classical) crystal. To see that an explanation at the individual pair interaction level is generally inadequate, consider Fig.~\ref{WU_NVT_NVP}(b) which shows a scatter plot of  single-particle energy and virial. These are sums over the pair interactions a given particle has with its neighbors; summing over all particles gives the total potential energy and virial, respectively. If strong correlation held at the level of single pair interactions, it would also hold at the particle level, but it clearly does not. This emphasizes that strong correlation is a collective effect, as detailed in Paper II.

In this paper we elaborate on the statistical mechanics and thermodynamics of strongly correlating liquids, and present results from computer simulations showing that strong virial /potential energy correlations are present even in non-equilibrium processes. The purpose is to present a number of new results supplementing those of Paper II in order to broadly illuminate the properties of the class of strongly correlating liquids. Together Papers II and III give a fairly complete characterization of the properties of a strongly correlating liquid at one state point, as well as at different state points with same density. Paper IV in this series\cite{IV} goes on to consider varying-density curves of ``isomorphic'' state points in the state diagram, which are characterized by several invariants; such curves exist only for strongly correlating liquids.

The organization of this paper is as follows. Section~\ref{IPL-ETC} begins with a discussion of the scaling properties of systems with inverse power-law (IPL) potentials, the natural starting point for a discussion of the hidden scaling properties of strongly correlating liquids. This is followed by a generalization to allow an extra term depending on volume only. Some, but not all, scaling properties of IPL systems are inherited by this generalization. Following this, in Sec. \ref{XIPL} we discuss the ``extended inverse-power law''(eIPL) potential introduced in Paper II, which includes the above-mentioned linear term. We illustrate with simulation results the key property that the linear term contributes significantly to the virial and potential energy fluctuations when the volume may fluctuate, but little when it is fixed. Hence it gives rise, approximately, to a volume-dependent term in the free energy of the type discussed in the previous subsection. This leads to an inherited approximate scaling property, which we refer to as ``hidden scale invariance'' since it is not immediately obvious from the intermolecular potential. The argument about how and why hidden scale invariance causes strong $WU$ correlations makes no assumption about equilibrium. To emphasize this point, Sec. \ref{SIM-RES} presents results from non-equilibrium computer simulations of strongly correlating molecular liquids, in particular aging following a temperature jump, and crystallization, both at constant volume. The property of strong correlation is shown to apply even in these out-of-equilibrium situations. Section~\ref{ENS-DEP} discusses ensemble-dependence, in particular it is here shown that the virial / potential energy correlation is always stronger in the NVT ensemble than in the NVE one. The last main section, Sec. \ref{THERMO}, comprises two topics under the heading ``thermodynamics of strongly correlating liquids''. First we discuss the relation of pressure-energy correlation to the thermodynamic Gr\"uneisen parameter $\gamma_G$, showing that the slope $\gamma$ (Eq.~(\ref{slopeDefinition})) is larger than $\gamma_G$ by roughly a factor involving the ratio of excess (configurational) to total specific heats (constant volume). This ratio is around two for many simple liquids.~\cite{Chisolm/Wallace:2001} The second part formulates the property of strong correlation in the energy-bond language known as ``network  thermodynamics''.\cite{Oster/Perelson/Katchalsky:1971}

\section{\label{IPL-ETC}Properties of inverse power-law systems and generalizations}

The purpose of this section is to summarize the properties of inverse-power law potentials and identify which of these properties are inherited by strongly correlating liquids and which are not.

\subsection{\label{IPL}Inverse power-law potentials}

Inverse power-law (IPL) potentials -- sometimes referred to as soft-sphere potentials -- have been used in liquid state theory for many years as convenient model systems.\cite{kle19,ber52,hoo70,hoo71,hiw74,benamotz03,dem04,ram09} Such potentials have a number of simple properties. IPL potentials have, however, been considered unrealistic because their predicted equation of state is quite wrong and because they have no stable low-pressure liquid phase and no van der Waals loop, problems which derive from the fact that IPL potentials are purely repulsive. Moreover, the correct IPL exponent fitting the Lennard-Jones (LJ) liquid is around $18$ (Papers I and II, Refs. \onlinecite{benamotz03,sti75,wee83}), not $12$ as one might naively guess from the repulsive $r^{-12}$ term of the LJ potential; this may have confused people searching from an effective IPL description of the LJ liquid. A major point  made in this series of papers is that when interpreted correctly, IPL potentials are much more realistic than generally thought, because they describe well a number of properties of strongly correlating liquids. For reference we now briefly summarize the since long well-established properties of IPL liquids.

Consider $N$ identical particles in volume $V$ interacting by a pair potential of the form $v(r)=Ar^{-n}$;  we make the pair assumption for simplicity but note that the below argument generalizes immediately to any potential that is an Euler homogeneous function of the position coordinates. From the standard partition function the (Helmholtz) free energy $F$ is conveniently written\cite{Hansen/McDonald:1986,Allen/Tildesley:1987}  as the ideal gas term plus the nontrivial ``excess'' free energy, $F=F_ {\rm id}+F_ {\rm ex}$. The first term is the free energy of an ideal gas at same volume and temperature, $F_ {\rm id}=-Nk_BT\ln(\rho\Lambda^3)$ where $\rho=N/V$ is the particle number density and $\Lambda=h/\sqrt{2\pi mk_BT}$ is the thermal de Broglie wavelength. The excess free energy is given\cite{Hansen/McDonald:1986,Allen/Tildesley:1987} by

\begin{equation}\label{fex}
e^{-F_ {\rm ex}/k_BT}\,=\,
\int \frac{d{\bf r}_1}{V}...\frac{d{\bf r}_N}{V}e ^{-U({\bf r}_1, ... , {\bf r}_N)/k_BT}\,.
\end{equation}
Whenever $n>3$ this expression leads to a free energy with a well-defined extensive thermodynamic limit ($N\rightarrow\infty$).\cite{kle19,ber52}

It follows from Eq. (\ref{fex}) that the excess free energy of an IPL liquid is given as follows in terms of a function of density $\rho$ to the power $n/3$ over temperature $T$, $ \phi(\rho^{n/3}/T)$ (Klein's theorem\cite{kle19,ber52}):

\begin{equation}\label{fex_ipl}
F_ {\rm ex, IPL}\,=\,
Nk_BT \phi\left(\rho^{n/3}/T\right)\,.
\end{equation}
This implies that a number of derived quantities are also functions of $\rho^{n/3}/T$. As important examples, recall the following standard identities: 
The excess entropy: $S_ {\rm ex}=-(\partial F_ {\rm ex}/\partial T)_V$, 
the potential energy: $U=F_ {\rm ex}+TS_ {\rm ex}$,
the virial $W=-V(\partial F_ {\rm ex}/\partial V)_T$,
the excess isothermal bulk modulus: $K_{T}^{\rm ex}=V(\partial^2 F_ {\rm ex}/\partial V^2)_T$,
the excess isochoric specific heat per unit volume: $c_{V}^{\rm ex}=-(T/V) (\partial^2 F_ {\rm ex}/\partial T^2)_V$,
the excess pressure coefficient: $\beta_{V}^{\rm ex}=(1/V) (\partial W/\partial T)_V=-\partial^2 F_ {\rm ex}/\partial T\partial V$.
From Eq. (\ref{fex_ipl}) it follows that these three quantities are functions of the single variable $\rho^{n/3}/T$; more accurately one has 
[where 
$f_1(x)=x\phi'(x)-\phi(x)$, 
$f_2(x)=x\phi'(x)$, 
$f_3(x)=(n/3)^2x^2\phi''(x)+[(n/3)+(n/3)^2]x\phi'(x)$, 
and $f_4(x)=-x^2\phi''(x)$]

\begin{eqnarray}\label{ipl_id}
S_ {\rm ex, IPL}\,&=&\,N k_B\, f_1 \left(\rho^{n/3}/T\right)\,, \\
U_{\rm IPL}\,&=&\, N k_BT \,f_2 \left( \rho^{n/3}/T\right) \,,\\
W_{\rm  IPL}\,&=&\frac{n}{3} Nk_BT\, f_2 \left(\rho^{n/3}/T\right) \,,\\
K_{T, \rm IPL}^{\rm ex}\,&=&   \rho k_BT\, f_3 \left(\rho^{n/3}/T\right)\,,\\
c_{V, \rm IPL}^{\rm ex} \,&=& \rho k_B f_4 \left(\rho^{n/3}/T\right)\,,\\
\beta_{V,\rm IPL}^{\rm ex}\,&=&\frac{n}{3}  \rho k_B\, f_4 \left(\rho^{n/3}/T\right)\,.
\end{eqnarray}
The functions $f_1, ..., f_4$ all depend on $n$, but for simplicity of notation we have not indicated this explicitly. Dividing across by the dimensional factors on the right hand side (for example $k_BT$ in the case of potential energy and virial), one arrives at dimensionless forms of the excess entropy, potential energy, etc. that are functions of $\rho^{n/3}/T$ only.

Turning now to the dynamics, consider the standard molecular dynamics (MD) case where the equations of motion are Newton's equations. Suppose ${\bf r}_i(t)$ ($i=1,...,N$) is a solution to Newton's equations. Then it is straightforward to show that ${\bf r}^{(1)}_i(t)=\alpha{\bf r}_i(\lambda t)$ is also a solution if $\alpha^{-(n+2)}=\lambda^2$. In particular, if ${\bf r}_i(t)$ refers to equilibrium (NVE or NVT) dynamics at a state point with density $\rho_0$ and temperature $T_0$, then ${\bf r}^{(1)}_i(t)=\alpha{\bf r}_i(\lambda t)$  refers to equilibrium dynamics at density $\rho_1=\rho_0/\alpha^3$ at temperature $T_1=T_0\alpha^2\lambda^2$ (temperature scales as the mean-square velocity and velocities get a factor $\alpha\lambda$). Using the  above relation between $\alpha$ and $\lambda$ this implies

\begin{equation}
T_1=\alpha^{-n}T_0= \left(\frac{\rho}{\rho_0}\right)^{n/3}T_0\,.
\end{equation}
This means that two states with different densities and temperatures but same $\rho^{n/3}/T$ have dynamics that scale into one another by simple scalings of space and time. In particular, if for any quantity $A$ one defines the relaxation time $\tau_{\rm A}$ via $\tau_{\rm A}=\int_0^\infty \langle A(0)A(t)\rangle dt/\langle A^2\rangle$, it follows that any two states with same $\rho^{n/3}/T$ have same "reduced" (dimensionless) relaxation time $\tilde\tau_{\rm A}$, if this quantity is defined by $\tilde\tau_{\rm A}=\tau_{\rm A}/t_0$ where the characteristic thermal time 
$t_0$ is defined by $t_0=\rho^{-1/3}\sqrt{m/k_BT}$. Similarly, if one defines the reduced diffusion constant $\tilde D$ by $\tilde D=D/D_{0}$ where 
$D_{0}=\rho^{-2/3}/t_{0}=\rho^{-1/3}\sqrt{k_BT/m}$, then $\tilde D$ is the same for the two states. Summarizing,

\begin{eqnarray}\label{dyn_id}
\tilde\tau\,&=&\,  f_5 \left(\rho^{n/3}/T\right)\,, \\
\tilde D\,&=&\,  f_6 \left(\rho^{n/3}/T\right)\,.
\end{eqnarray}
Finally we note that it follows from the above scaling property that  

\begin{eqnarray}
 \frac{U_{IPL}({\bf r}^{(1)}_1, ... , {\bf r}^{(1)}_N)}{k_BT_1}\, =\, \frac{U_{IPL}({\bf r}_1, ... , {\bf r}_N)}{k_BT_0}\,.
\end{eqnarray}
Thus the Boltzmann factors of the two configurations are the same and, consequently, the scaling of the dynamics holds also for stochastic dynamics; this observation, in a generalized form, is the starting point of Paper IV in this series. By the same argument the structure of states with same $\rho^{n/3}/T$ are identical, provided lengths are scaled by $\rho^{-1/3}$.

\subsection{\label{INHERIT}Inheritance of scaling properties by generalized IPL potentials}

This section discusses how a large class of potentials inherits a number of IPL properties to a good approximation, thus justifying the term ``hidden scale invariance''.  Consider a general potential between particles $i$ and $j$, rewriting it (as can always be done) as a sum of an IPL potential plus the difference, denoted ``diff'':

\begin{eqnarray}
   v_{ij}(r_{ij}) = \epsilon^{}_{ij}\left(\frac{\sigma^{}_{ij}}{r_{ij}}\right)^{n} + v^{\rm diff}_{ij}(r_{ij})\,.
\end{eqnarray}
For any configuration, $( {\bf r}_1, ... , {\bf r}_N)$, the potential energy is then the sum of an ``IPL'' term and a ``diff'' term, and the excess free energy is given by

\begin{equation}
e^{- F_ {\rm ex}/k_BT}\,=\,
\int \frac{d{\bf r}_1}{V}...\frac{d{\bf r}_N}{V}
   e ^{-U_{\rm IPL}({\bf r}_1, ... , {\bf r}_N)/k_BT}
   e ^{-U_{\rm  diff}({\bf r}_1, ... , {\bf r}_N)/k_BT}\,.
\end{equation}
We now investigate consequences of the assumption (Sec. III) that $U_{\rm diff}$ to a good approximation is only a function of volume:  
$U_{\rm diff}({\bf r}_1, ... , {\bf r}_N) \cong f(V)$ -- at least for states that carry Boltzmann weights of any significance. The approximate identity 
$U_{\rm diff}({\bf r}_1, ... , {\bf r}_N) = f(V)$ means that the second exponential  can be moved outside the integral, and we get: 

\begin{equation}
e^{-F_ {\rm ex}/k_BT}\,=\, 
e^{-f(V)/k_BT}
\int \frac{d{\bf r}_1}{V}...\frac{d{\bf r}_N}{V}
   e ^{-U_{\rm IPL}({\bf r}_1, ... , {\bf r}_N)/k_BT} \,.
\end{equation}
From this follows directly that

\begin{eqnarray}\label{inherit_f_id}
F_{\rm ex} &=& f(V) + F_{\rm ex, IPL} =  f(V) + Nk_BT \phi \left(\rho^{n/3}/T\right)\,,
\end{eqnarray}
which implies

\begin{eqnarray}\label{inherit_ipl_id}
S_{\rm ex}\,&=&\,S_{\rm ex, IPL} = N k_B\, f_1 \left(\rho^{n/3}/T\right)\,, \\
U &=& f(V) + U_{\rm IPL}\,=\,f(V) + N k_BT \,f_2 \left(\rho^{n/3}/T\right) \,,\\
W &=& - f'(V)V + W_{\rm  IPL}\,=  - f'(V)V +\frac{n}{3} N k_BT \,f_2 \left(\rho^{n/3}/T\right)\,,\\
K_{T}^{\rm ex} &=& Vf''(V)+K_{T,\rm IPL}^{\rm ex}\,=\,Vf''(V)+\rho k_BT\, f_3 \left(\rho^{n/3}/T\right)\,,\\
c_{V}^{\rm ex} \,&=& c_{V,\rm IPL}^{\rm ex} = \rho k_B f_4 \left(\rho^{n/3}/T\right)\,,\\
\beta_{V}^{\rm ex}\,&=&\beta_{V,\rm IPL}^{\rm ex} = \frac{n}{3}\rho k_B\, f_4 \left(\rho^{n/3}/T\right)\,.
\end{eqnarray}

While the systems under consideration here have the same excess entropy as the ``hidden'' IPL, several quantities have contributions from the $f(V)$-term. These quantities do not obey IPL scaling. In contrast, the scaling behavior for dynamics and structure \emph{is} inherited: Consider two state points with the same  $\rho^{n/3}/T$. For the pure IPL system ($f(V)=0$) the two state points have the same dynamics and structure as argued in the previous section. Letting $f(V)\neq 0$ simply shifts the energy surface, which changes neither the dynamics nor the structure.\cite{Schroder/others:2008,Schroder/Pedersen/Dyre:2008a} This scenario -- scaling of the dynamics, but not the equation of state -- is exactly what is observed experimentally for a large number of viscous liquids. For example in van der Waals liquids relaxation times are found to be a function of $\rho^{n/3}/T$ (using $n$ as an empirical parameter),\cite{Alba-Simionesco/others:2004,Roland/others:2005} but the scaling does not apply to the (excess) pressure with the exponent determined from the scaling of relaxation time, as required for IPL scaling.\cite{Alba-Simionesco/others:2004,Grzybowski/Paluch/Grzybowska:2009}

In the following section we provide numerical evidence that there are indeed systems that to a good approximation fulfil the assumption introduced above that $U_{\rm diff}$ is a function of volume only.

\section{\label{XIPL}Lennard-Jones as a generalized IPL potential: the extended inverse power-law potential approximation}

In this section we examine the extent to which the Lennard-Jones (LJ) potential may be approximated by an ``extended inverse power-law'' potential (including a linear term, Eq. (\ref{v_eIPL}) below), by considering the 
fluctuations at a particular state point of the LJ fluid. We choose a state point whose pressure is near zero, because here it is particularly clear that single-pair effects are insufficient to explain the strong $WU$ correlation (Fig. 1). 

The analysis of Paper II took its starting point in assuming that the approximating inverse power law should match the potential closely at a particular value of interparticle separation. An important conclusion of the analysis was, however, that the success of the IPL approximation derives not from its behavior near any particular $r$-value, but rather from the fact the difference from the real potential is close to linear over the whole first-peak region. In this section we re-examine the idea that the fluctuations are well described by an inverse power-law potential and the argument for why the difference term almost does not fluctuate. We show explicitly that the latter contributes little to the fluctuations at constant volume, but significantly when the volume is allowed to fluctuate as in the NpT ensemble. This demonstrates that the LJ potential is of the type considered in the previous subsection.

We wish to determine to what extent the LJ potential

\begin{equation}
v_{LJ}(r) = 4{\epsilon}\left(  (\sigma/r)^{12} -(\sigma/r)^6 \right)
\end{equation}
can be matched, for the purpose of describing fluctuations of potential energy and virial at fixed volume, by an inverse power law

\begin{equation}
v_{IPL}(r) = A r ^{-n} [+B]\,.
\end{equation}
Here $B$ indicates an optional constant. To start with, how should the exponent $n$ and the coefficients $A$ and $B$ be chosen? An obvious choice, followed in the first part of Paper II, is to require that the two potentials, $v_{LJ}$ and $v_{IPL}$, should agree as much as possible around a particular value of $r$, denoted $r_0$. Given $r_0$, if we require the functions and their first two derivatives to match at $r_0$, this determines all three parameters $A$, $n$ and $B$. The exponent $n$ is given (Paper II) by 

\begin{equation}
n = n^{(1)}(r_0) \equiv -\frac{r_0 v_{LJ}''(r_0)}{v_{LJ}'(r_0)} - 1\,.
\end{equation}

\nod Here the notation $n^{(1)}(r_0)$ refers to one kind of $r$-dependent effective inverse power-law exponent, based on the ratio of the second and first derivatives. For $v_{IPL}(r)$ this simply returns $n$. Otherwise it gives a local matching of the $v_{IPL}(r)$ to $v_{LJ}(r)$. This leaves effectively one parameter to vary, namely $r_0$, which must be less than the minimum $r_m=2^{1/6}\sigma$ where $n^{(1)}$ diverges. The parameter $r_0$ may be chosen to optimize the match of the fluctuations in total energy and virial. For an NVT simulation at $T=80$K and near-zero pressure, the best fit was obtained with $n=19.2$ (while the exponent implied by the slope [Eq. (\ref{slopeDefinition})], $\gamma=6.3$, was slightly smaller, 18.9). 

Later in Paper II  it was demonstrated that there is no particular reason why the potentials should match close at a particular value of $r$, since the fluctuations have contributions from the whole first-peak region, including beyond the potential minimum. The reason that any kind of matching is possible over this region -- where $v_{LJ}$ clearly does not resemble a power law -- is that a linear term may be added to the power-law potential almost without affecting the fluctuations {\em as long as the volume is held constant}. The analysis of Paper II, which also included an in-depth treatment of the perfect LJ (fcc) crystal which is also strongly correlating, showed that the more relevant $r$-dependent effective exponent is the higher order $n^{(2)}$ defined by

\begin{equation}
n^{(2)}(r_0) \equiv -\frac{r_0 v_{LJ}'''(r_0)}{v_{LJ}''(r_0)} - 2\,.
\end{equation}
This also returns $n$ for $v_{IPL}(r)$, but since it does not involve the first derivative, it returns $n$ even if a linear function of $r$ is added to the potential as in the extended inverse power-law potential (eIPL) defined by

\begin{equation}\label{v_eIPL}
v_{eIPL} =  A r^{-n} + B + Cr\,.
\end{equation}
This potential fits the LJ potential very well around its minimum (Paper II) and thus includes part of its attractive part.

There are several possible ways of choosing the ``best'' eIPL to match the real potential. These will give slightly different exponents and coefficients $A$, $B$ and $C$. We do not investigate them here; rather the purpose is to validate the basic idea of the extended inverse power-law (eIPL) approximation. Therefore we choose a simple matching scheme, whereby we match the fluctuations to those of the inverse power-law potential, without including a linear term, in order to determine $n$ and $A$. For simplicity we take the exponent directly from the observed fluctuations: $n=3\gamma$ where $\gamma$ is defined in Eq.~(\ref{slopeDefinition}). To fix the coefficient $A$, agreement with the potential energy and virial fluctuations is optimized by proceeding as follows: For a given configuration generated in an LJ molecular dynamics simulation we calculate the LJ potential energy $U_{LJ}$ and the power-law potential energy $U_{IPL}$, similarly the corresponding virials $W_{LJ}$  and $W_{IPL}$. The difference quantities $U_{\textrm{diff}}$ and $W_{\textrm{diff}}$ are defined as

\begin{align}
U_{\textrm{diff}} &= U_{LJ} - U_{IPL}\,, \\
W_{\textrm{diff}} &= W_{LJ} - W_{IPL}\,.
\end{align}
A perfect match of the fluctuations would mean that $U_{\textrm{diff}}$ and $W_{\textrm{diff}}$ have zero variance. Therefore we choose $A$ to 
minimize the sum of the relative ``diff'' variances:

\begin{equation}\label{match}
\frac{\angleb{(\Delta U_{\textrm{diff}})^2}}{\angleb{(\Delta U_{LJ})^2}} +
\frac{\angleb{(\Delta W_{\textrm{diff}})^2}}{\angleb{(\Delta W_{LJ})^2}}\,.
\end{equation}

\begin{figure}
\includegraphics[width=10cm]{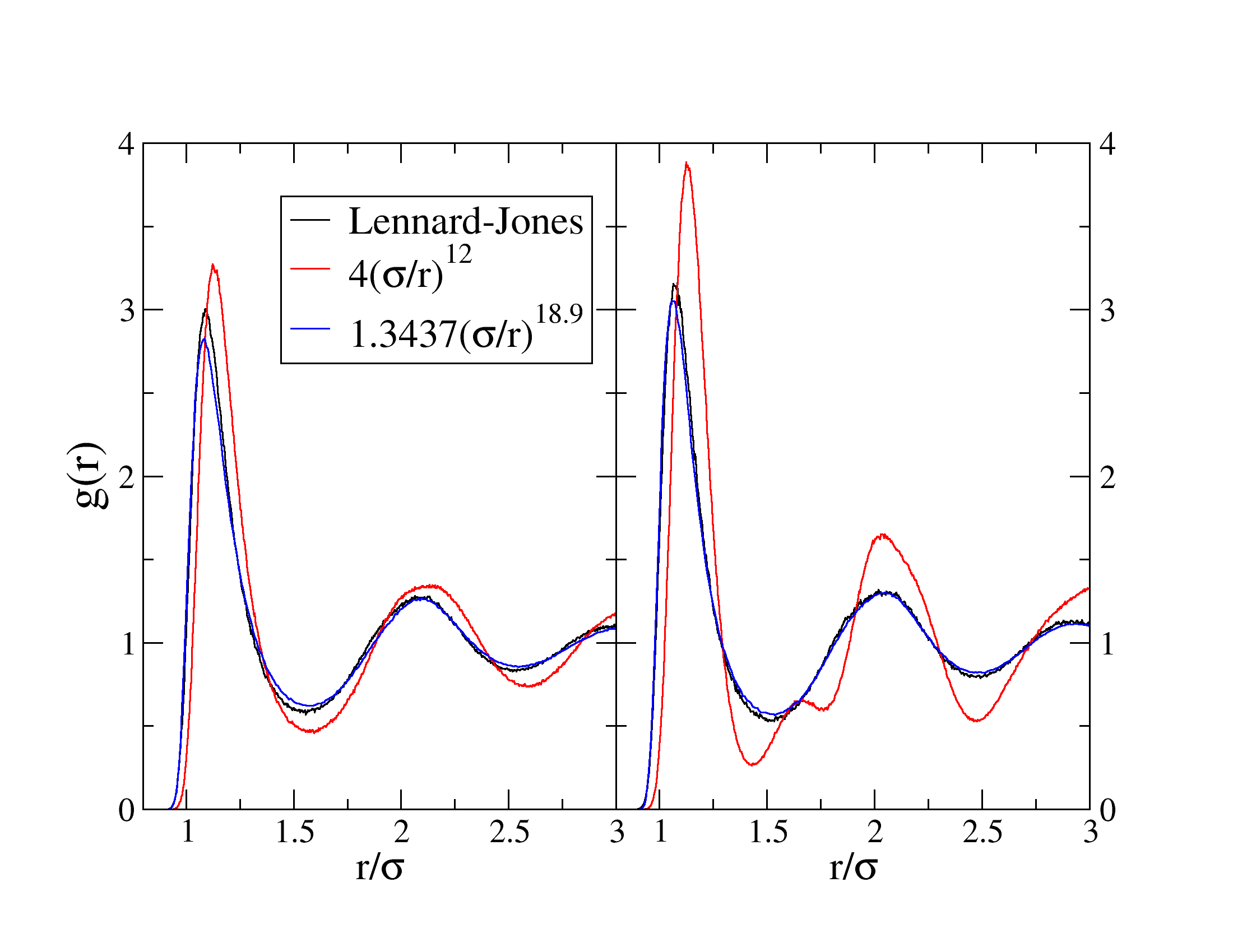}
\caption{\label{compareGR} (Color) Comparison of $g(r)$ for simulations using the Lennard-Jones potential and two inverse power-law potentials: the $r^{-12}$ repulsive term in $v_{LJ}(r)$ and the inverse power-law potential that optimizes the agreement in the fluctuations of potential energy and virial by minimizing Eq. (\ref{match}). The left panel shows these at density 0.82 and temperature 0.67 (dimensionless units), the right one at density 0.90 and temperature 0.80 (where the $r^{-12}$ potential leads to crystallization). }
\end{figure}

For the near-zero pressure state point used in Fig.~1 of Paper I the exponent determined from $\gamma$ is $n=3\gamma =18.9$ and the optimal value of $A$ is 1.3437$\epsilon \sigma^n$. Before examining the difference potential, what do we get if we simulate with the matched inverse power-law potential? Figure~\ref{compareGR} shows the radial distributions $g(r)$ obtained for the above state point and another with a higher density and temperature, for three potentials: LJ, the repulsive $r^{-12}$ term of the LJ potential, and the optimal IPL potential with $n=18.9$. We used the same inverse power-law potential at both state points (i.e., we did not adjust $A$ and $n$ to match the second state point). The first thing to note is that the $n=18.9$ potential gives a structure much closer to that of LJ than does the repulsive $n=12$ term alone, in particular the latter system has crystallized at the higher density and temperature. The second point is that there is still a difference between the LJ and the $n=18.9$ IPL, present in both state points. The first peak in the LJ system is slightly higher and narrower, although its position is barely altered. Thus the real potential gives a slight increase of order -- the difference in coordination number is less than 0.1 (integrating to the first minimum after the peak).

\begin{figure}
\includegraphics[width=10cm]{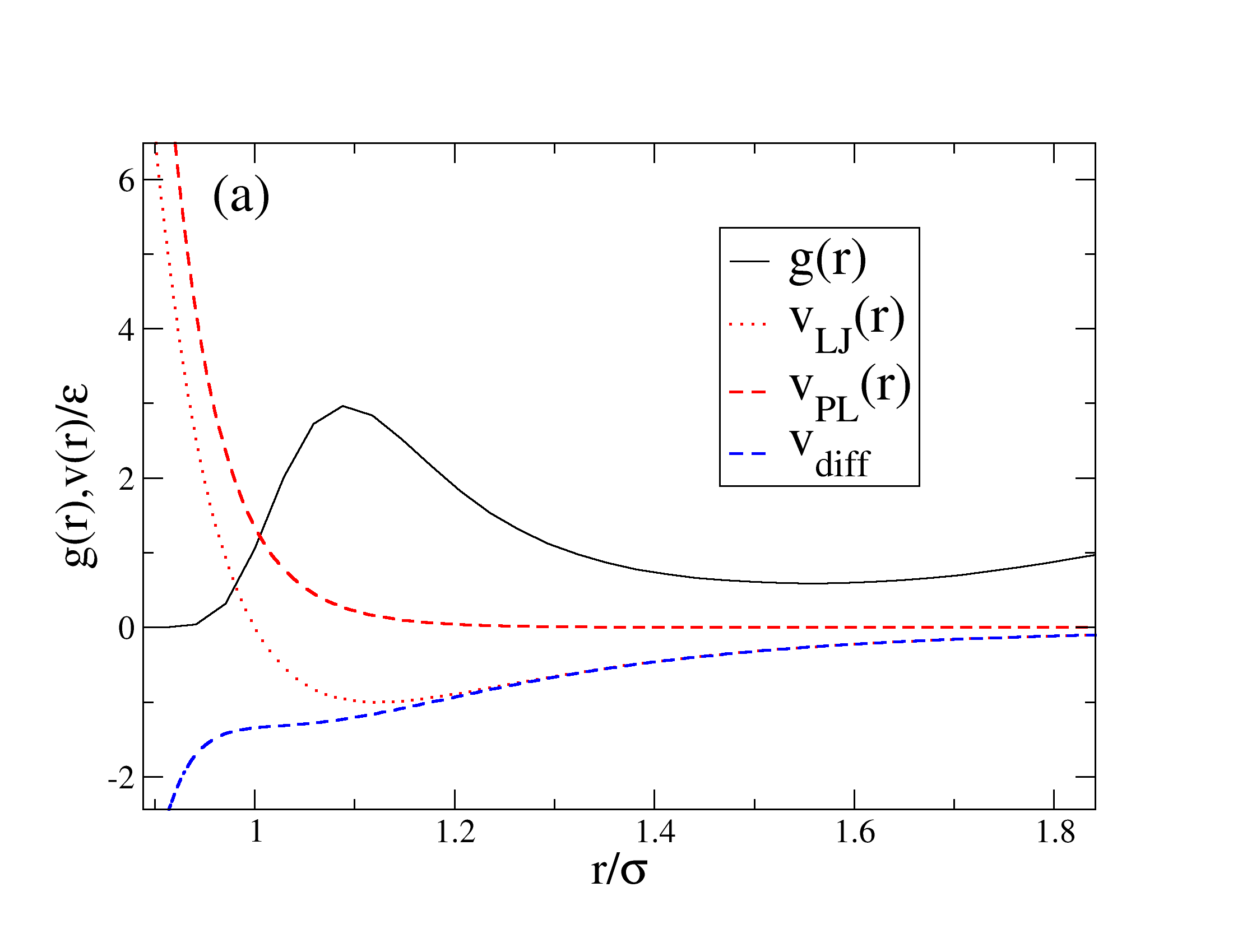}
\includegraphics[width=10cm]{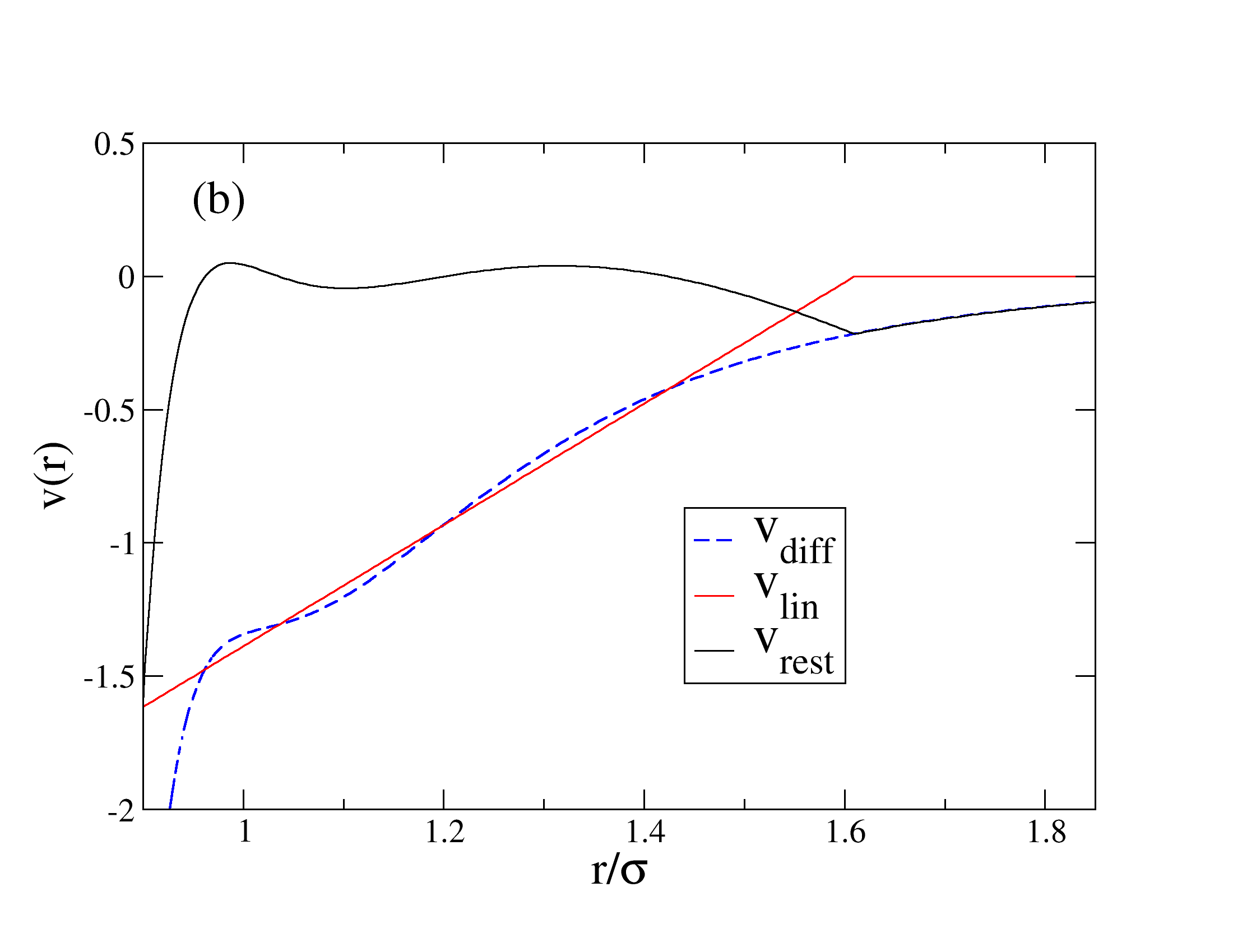}
\caption{\label{empiricalPLmatch} 
(Color)
(a) Illustration of the difference between the Lennard-Jones (LJ) potential $v_{LJ}(r)$, the empirically matched inverse power-law potential $v_{IPL}(r)$ with $A=1.3437\epsilon\sigma^n$ and $n=18.9$, and their difference $v_{\textrm{diff}}(r)$.  
(b) Linear fit, $v_{\textrm{lin}}(r) = \textrm{min}(0,-3.6635+2.2756r/\sigma)$, to $v_{\textrm{diff}}$ between 0.95$\sigma$ and 1.5$\sigma$, and the remainder $v_{\textrm{rest}}(r) \equiv 
v_{\textrm{diff}}(r) - v_{\textrm{lin}}(r)$ (full black curve).}
\end{figure}

\begin{figure}
\includegraphics[width=10cm]{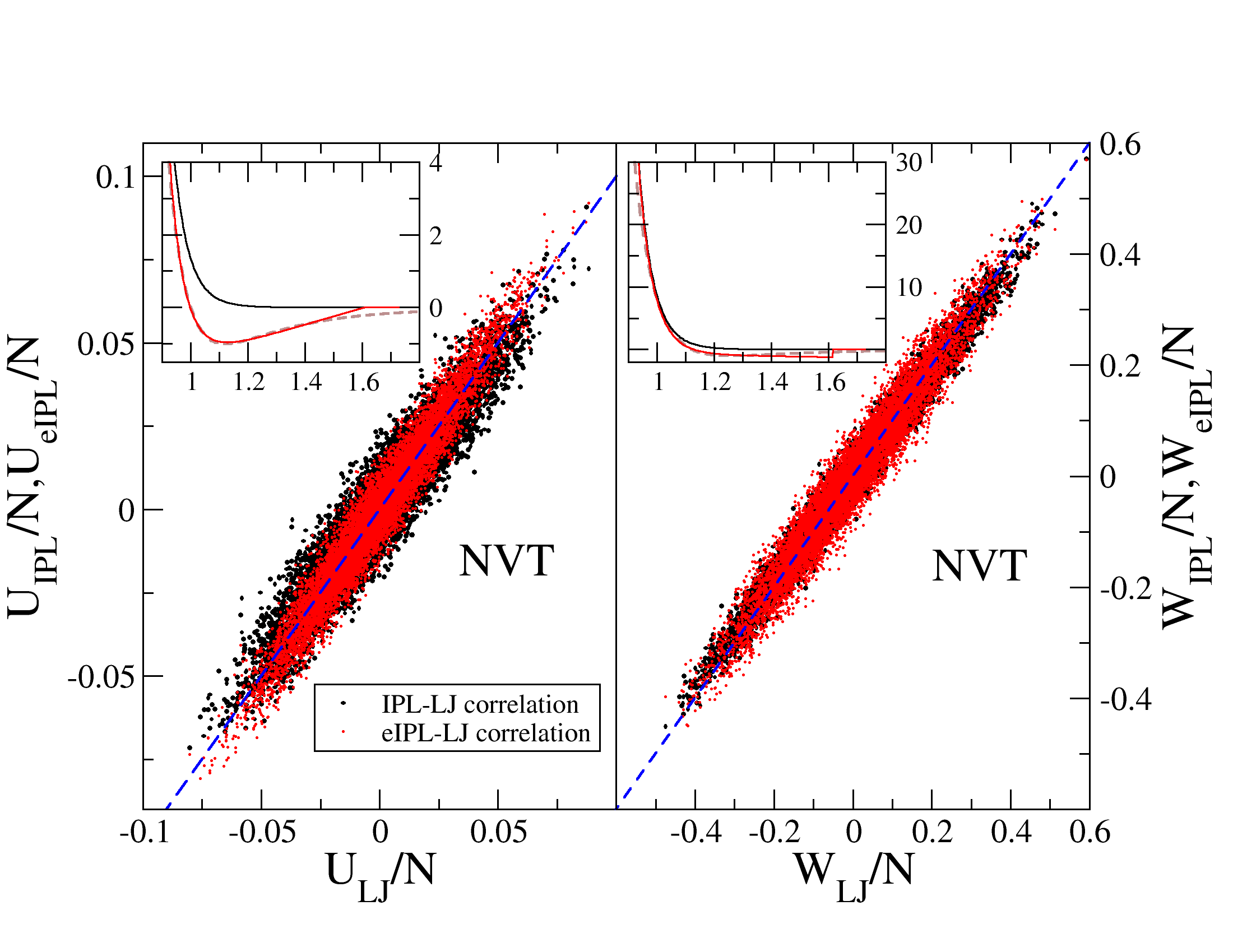}
\caption{\label{showCompLJ_PL_XIPL_NVT} (Color) 
Effect on fixed-volume fluctuations of adding a linear term to the inverse power-law (IPL) potential. The linear term is that shown in Fig.~\ref{empiricalPLmatch}(b). Configurations were generated by an NVT simulation using the LJ potential, and the different determinations of energy (LJ, IPL and eIPL) and virial were computed on these configurations. The dashed lines indicate a perfect match. Including the linear term when computing the energy improves the match to the true (LJ) fluctuations (the correlation coefficient goes from 0.950 to 0.970), while it reduces the match to the virial (the correlation coefficient goes from 0.987 to 0.971, which is probably related to the fact that the pair virial is discontinuous at $r_c$ -- we thus find that smoothing the linear part around $r_c$ restores the match somewhat). The insets show the pair potentials and virials: Brown dashed lines: LJ, black lines: IPL, red lines: eIPL. -- The overall conclusion from Fig. \ref{showCompLJ_PL_XIPL_NVT} is that the addition of the linear term induces little change in the fluctuations.}
\end{figure}

Figure~\ref{empiricalPLmatch}(a) shows the LJ potential, the IPL potential with parameters optimized as described above, their difference, and the radial distribution function. As shown in Fig.~\ref{empiricalPLmatch}(b), the main part of the difference potential $v_{\textrm{diff}}(r)$ is nearly linear. Thus a good approximation to the real potential is the eIPL potential of Eq.~(\ref{v_eIPL}) for $r$ less than a cut-off $r_c$, and zero otherwise. Neglecting the small value of $v_{IPL}(r_c)\sim 10^{-5}\epsilon$, the cut-off is given by $r_c = -B/C$.  For the fit shown in Fig.~\ref{empiricalPLmatch} (b), $r_c=1.61\sigma$. 

What are the implications of the linear term?  A linear term in the pair potential contributes a term proportional to the sum of all bond lengths to the total potential energy.  It was shown in Paper II that at constant volume this sum is a constant in one dimension, and it was argued that it is approximately constant in three dimensions. The difference is because of two things: First, in three dimensions there are contributions to bond-length changes from transverse components of relative displacements between the two particles defining a bond; secondly, within the eIPL the potential is only linear up to $r_c$ and, moreover, significant deviations of $v_{\textrm{diff}}$ from linearity occur at $r$'s smaller than $r_c$).

The argument that the linear term contributes little to the fluctuations depends on all bond lengths being less than $r_c$. In one dimension, at moderate temperatures,
a single-component system has a rather well-defined nearest-neighbor distance, which at densities where the pressure is not too negative will be less than $r_c$. In a three-dimensional liquid, however, the nearest-neighbor distance is not as well-defined -- the radial distribution function does not go to zero after the first peak. Therefore there will always be fluctuations at $r_c$ as the lengths of bonds fluctuate back and forth across $r_c$, so the sum of bond lengths which are less than $r_c$ will fluctuate. In Paper II it was shown that for a three-dimensional (classical) crystal at low temperature -- where this is not an issue because $g(r)$ does go to zero after the peak -- the correlation coefficient $R$ becomes very high, over 99.5\%, as $T\rightarrow0$ (but not 100\%).

\begin{table}
\caption{\label{WUbreakdown}Variances of potential energy $U$ and virial $W$, and of various contributions to $U$ and $W$, of two different ensembles at the LJ state point given by $\rho=0.82$ and $T=0.67$ (dimensionless units).}
\begin{tabular}{|c|c|c|c|c|c|c|}
\hline
Ensemble & Quantity & LJ & IPL & diff & lin & rest \\
\hline
\hline
NVT & $U$ & 0.0231 & 0.0225 & 0.0075 & 0.0085 & 0.0063 \\
& $W$ & 0.1468 & 0.1402 & 0.0227 & 0.0301 & 0.0350 \\
\hline
NpT & $U$ & 0.0484 & 0.0320 & 0.0665 & 0.0539 & 0.0144 \\
& $W$ & 0.1704 & 0.1997 & 0.0589 & 0.0417 & 0.0430 \\
\hline
\end{tabular}
\end{table}

We can check directly the effect of adding the linear term to the inverse power law. Figure~\ref{showCompLJ_PL_XIPL_NVT} shows scatter plots of IPL (black) and eIPL (red) energy and virial plotted against the true LJ values. These were calculated for a set of configurations drawn from an NVT simulation using the true (LJ) potential. Including the linear term makes little difference. It somewhat improves the match to the energies, though not to the virials (possibly due to the discontinuity in the pair-virial at $r_c$). The $WU$ correlation coefficient of the eIPL potential energy and virial is 0.917 (compare to the true (LJ) value of 0.938 and the pure IPL value of 1.0).

\begin{figure}
\includegraphics[width=10cm]{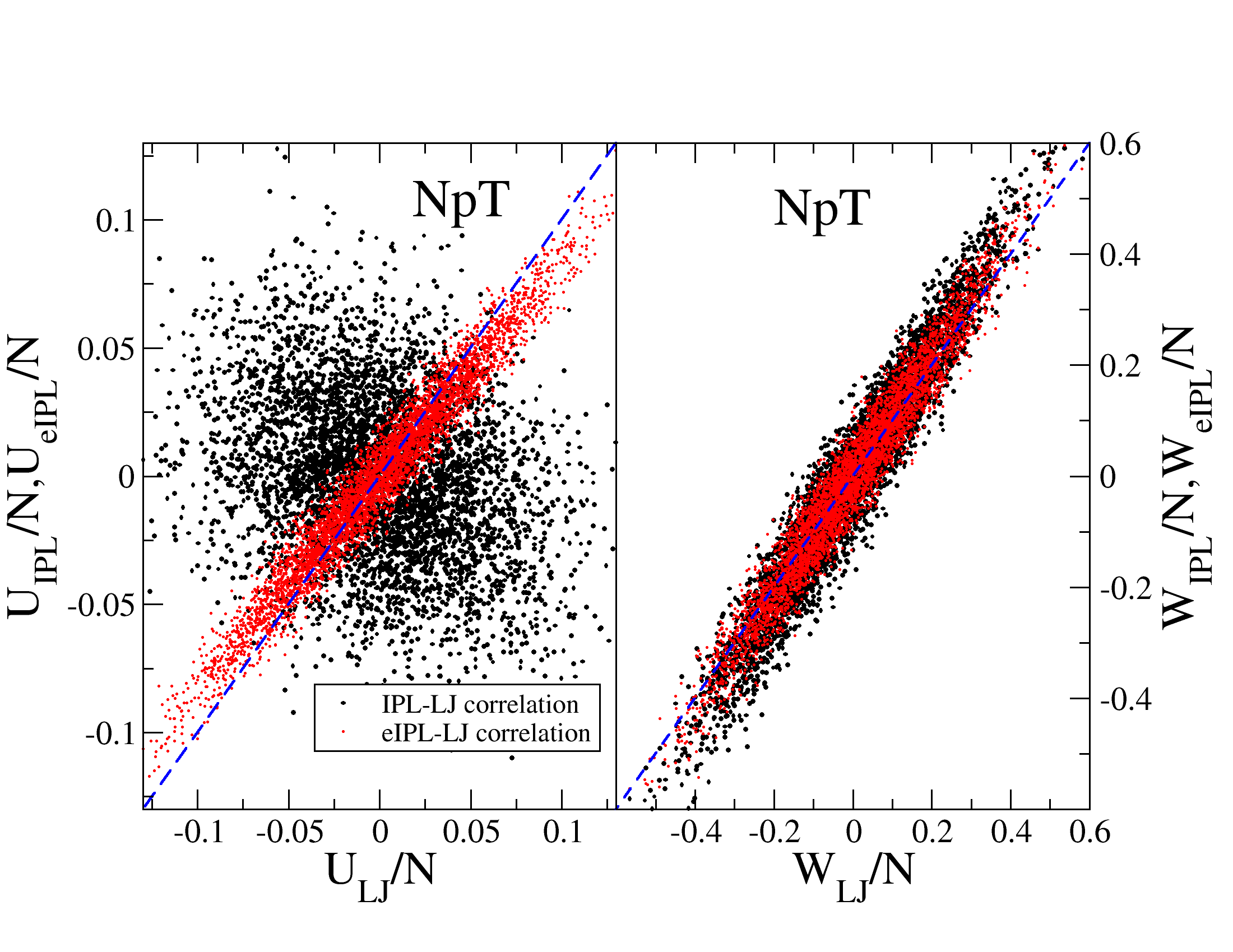}
\caption{\label{showCompLJ_PL_XIPL_NPT} (Color) Comparison of potential energy calculated using $v_{LJ}(r)$, $v_{IPL}(r)$ (black points) and $v_{eIPL}(r)$ (red points) for configurations drawn from an NpT simulation using the LJ potential. The potential energy, in particular, is very poorly represented by the power-law contribution when the volume is allowed to fluctuate. In fact the correlation between $U_{IPL}$ and $U_{LJ}$ is not only weak, it is negative. Including the linear term makes a huge difference here, yielding a correlation coefficient of 0.977 between $U_{eIPL}$ and $U_{LJ}$ (changed from -0.201). The slope is somewhat less than unity, indicating that there are significant contributions from pair distances beyond $r_c$ (i.e., from the ``rest'' part of the potential). The linear term affects the virial fluctuations much less, presumably because the derivative of the potential is dominated by the IPL term.}
\end{figure}

It is instructive to repeat the above for configurations drawn from an NpT simulation at the same state point, i.e., with pressure chosen as the average pressure in the NVT case. The results are shown in Fig.~\ref{showCompLJ_PL_XIPL_NPT}. Here it is clear that the IPL potential represents the potential energy fluctuations very poorly (black points), while adding the linear term makes a substantial difference (red points). As in the NVT case the linear term affects the virial fluctuations much less. This is presumably because when taking the derivative to form the virial, the IPL term gets  multiplied by $n=18$ while the linear term gets multiplied by minus one, and is thus reduced considerably in significance (compare the insets of Fig.~\ref{showCompLJ_PL_XIPL_NVT}).

The size of the variances of the different terms are compared in Table~\ref{WUbreakdown}. We do not make a detailed analysis of the variance (taking into account cross-correlations, etc). In the NVT case the IPL contributions are of similar size to the full (LJ) fluctuations -- naturally since we explicitly optimized this --  and the ``diff'' contributions are small compared to the IPL ones. In the NpT case, on the other hand, the ``diff'' contributions to the fluctuations of $U$ are more than double the IPL ones; this is not the case for the ``diff'' contributions to $W$, though they are still a larger fraction of the total than in the NVT case. These numbers are consistent with Fig.~\ref{showCompLJ_PL_XIPL_NPT}: The ``diff'' contributions to the energy must be larger than the IPL ones because the latter are negatively correlated with the true energy. The fact that the variance of $U_{\textrm{diff}}$ is smaller than the sum of those of $U_{\textrm{lin}}$ and $U_{\textrm{rest}}$, in the NVT case, indicates that the latter are negatively correlated. This is presumably due to bond-lengths around $r_c$ which alternately are counted as part of $U_{\textrm{lin}}$ and as part of $U_{\textrm{rest}}$. This effect is less noticeable in the NpT case; there we see clearly that fluctuations in $U_{\textrm{lin}}$ account for most of those in $U_{\textrm{rest}}$.

Based on the above we can now answer the question: Is it possible to predict whether or not a liquid is strongly correlating by inspection of its potential (i.e., without simulating virial and potential energy fluctuations)? For liquids with particles interacting by pair potentials, the answer is the affirmative: The liquid is strongly correlating if the potential around the first peak of the structure factor (the typical interparticle distance) may be fitted well by an extended inverse power law. For more general potentials, the situation is more complex. Thus it is possible to construct many-body potentials with angular dependencies which scale like inverse power-law pair potentials; these have 100\% $WU$ correlation because this property follows whenever the potential is an Euler homogeneous function. In most realistic cases, however, systems with angular dependencies are not expected to be strongly correlating; likewise potentials with two length scales will generally not be strongly correlating. An example of the former is the coarse-grained model of water using a short-range many-body potential recently introduced by Molinero and Moore \cite{mol09} that reproduces water's properties with surprising accuracy. This potential is not strongly correlating, because close to water's density maximum the virial / potential energy correlation coefficient $R$ must be close to zero (Paper I). Examples of potentials with two length scales, that for this reason are not strongly correlating, are the Jagla potential\cite{jagla} and the Dzugutov potential.\cite{Dzugutov:1992} Likewise, the addition of Coulomb terms to an LJ-type liquid generally ruins strong correlations (Paper I).

\section{\label{SIM-RES}Out-of-equilibrium dynamics in molecular models}

\begin{figure}
\begin{center}
\includegraphics[width=6cm]{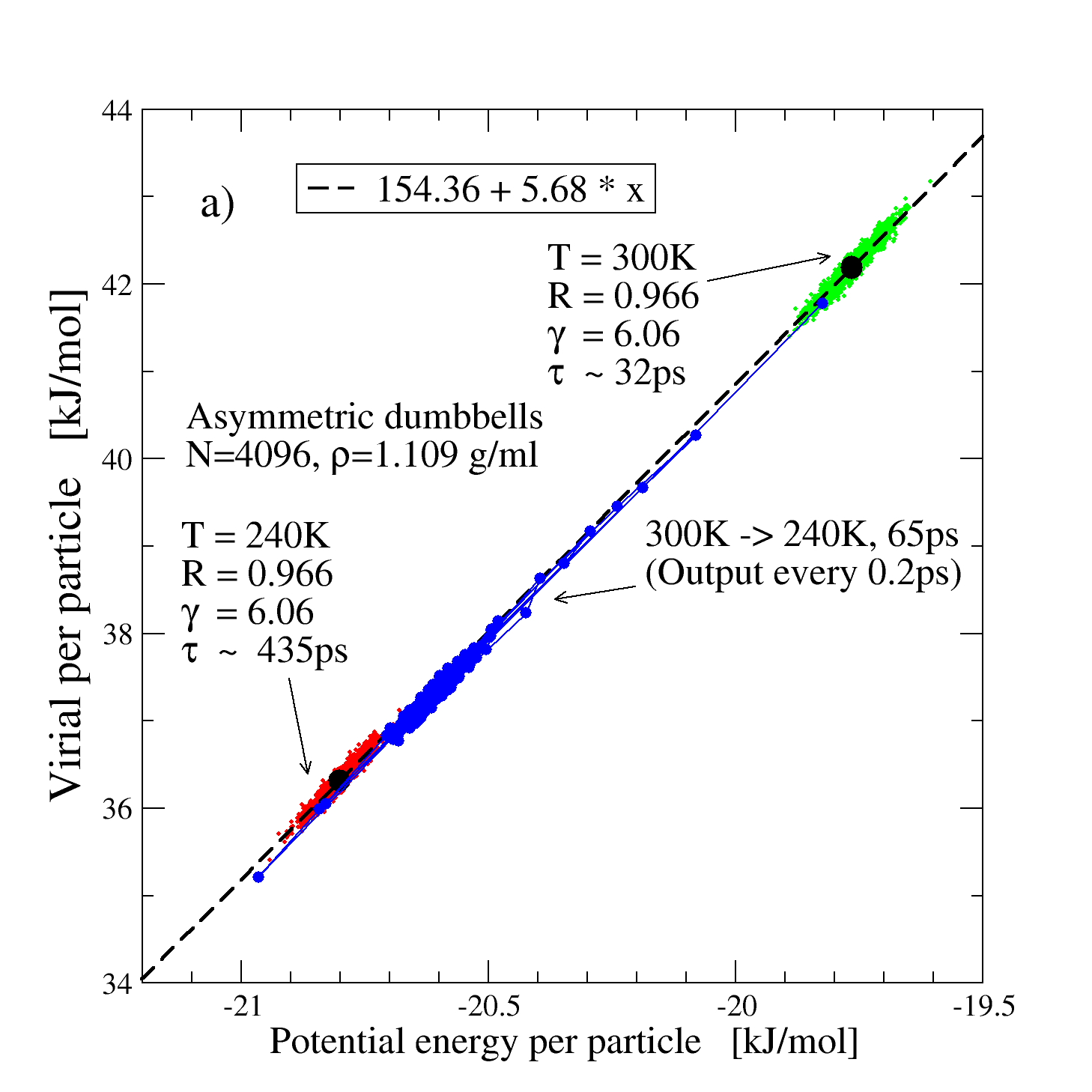}
\includegraphics[width=6cm]{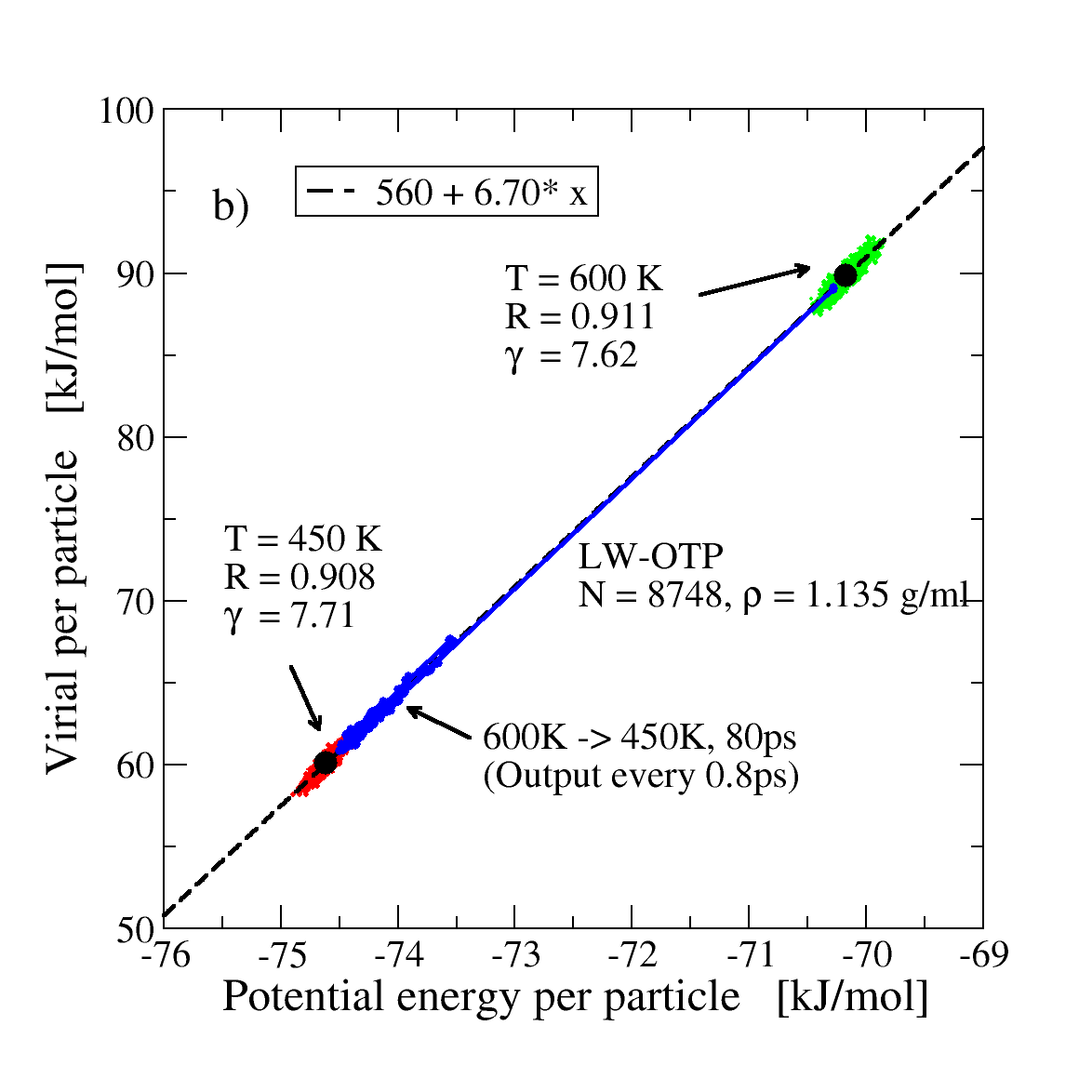}
\includegraphics[width=6cm]{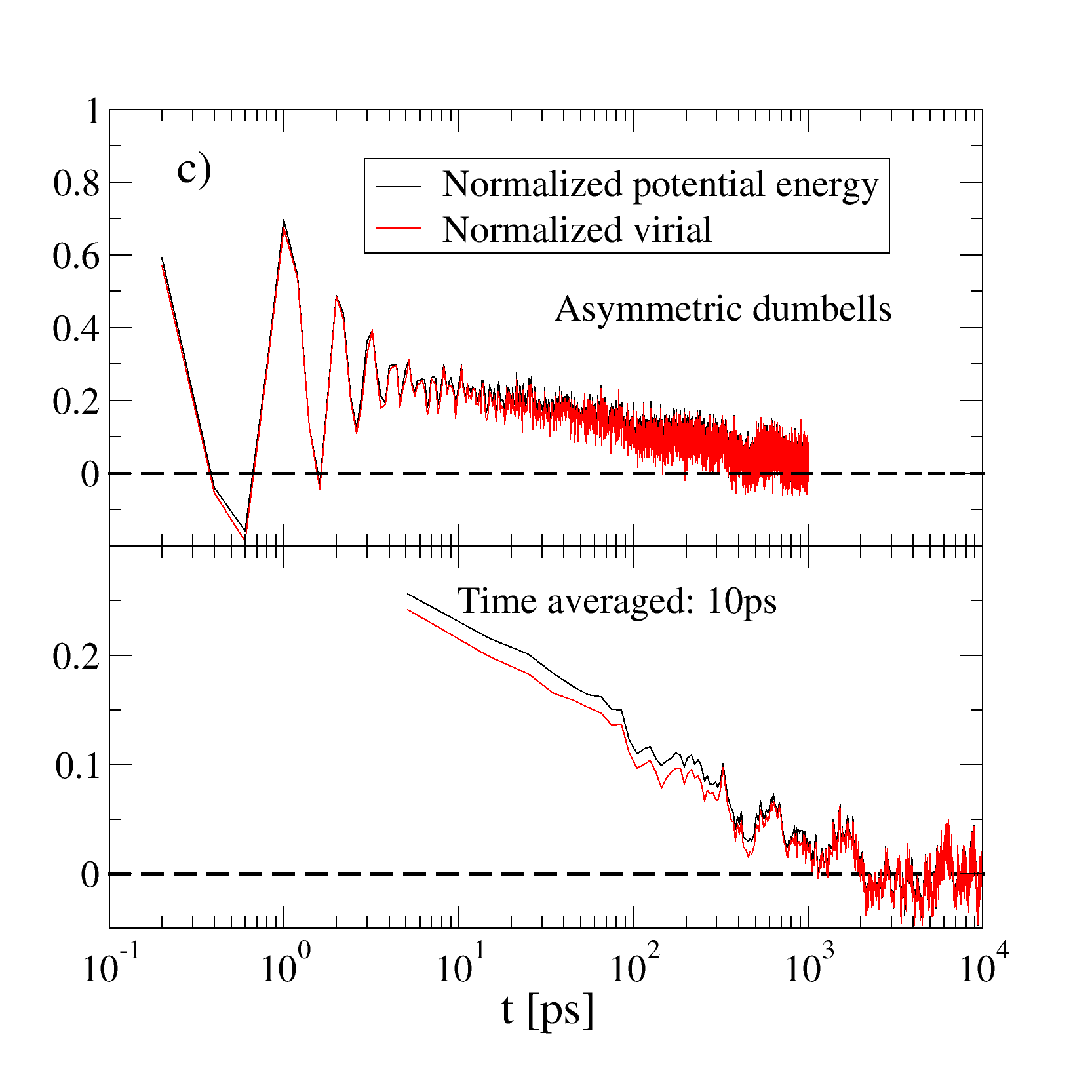}
\end{center}
\caption{
(Color) Computer simulations of virial and potential energy during the aging of two strongly correlating molecular liquids following temperature down-jumps at constant volume (NVT simulations). 
(a) The asymmetric dumbbell model at density $\rho = 1.109$ g/ml. The liquid was first equilibrated at T=300 K. Here simultaneous values of virial and potential energy are plotted at several times producing the green ellipse, the elongation of which directly reflects the strong $WU$ correlation in equilibrium. Temperature was then changed to T=240 K where the red ellipse marks the equilibrium fluctuations. The aging process itself is given by the blue points. These points follow the line defined by the two equilibrium simulations, showing that virial and potential energy correlate also out of equilibrium. 
(b) Similar temperature down-jump simulation of the Lewis-Wahnstr{\"o}m OTP system.\cite{Lewis/Wahnstrom:1994} The colors have the same meaning as in (a): Green marks the high-temperature equilibrium (T=600 K), red the low-temperature equilibrium (T=450 K), and blue the aging towards equilibrium. 
--  In both (a) and (b) the slope of the dashed line is not precisely the number $\gamma$ of Eq. (\ref{slopeDefinition}) because the  liquids are not perfectly correlating; the line slope is $\langle\Delta U\Delta W\rangle/\langle(\Delta U)^2\rangle$, see Paper I, a number that is close to $\gamma$ whenever the liquid is strongly correlating.
(c) Virial and potential energy for the asymmetric dumbbell model as functions of time after the temperature jump of (a); in the lower subfigure data were averaged over 10 ps. Virial and potential energy clearly correlate closely.}
\label{WU_aging_scl}
\end{figure}

\begin{figure}
\begin{center}
\includegraphics[width=10cm]{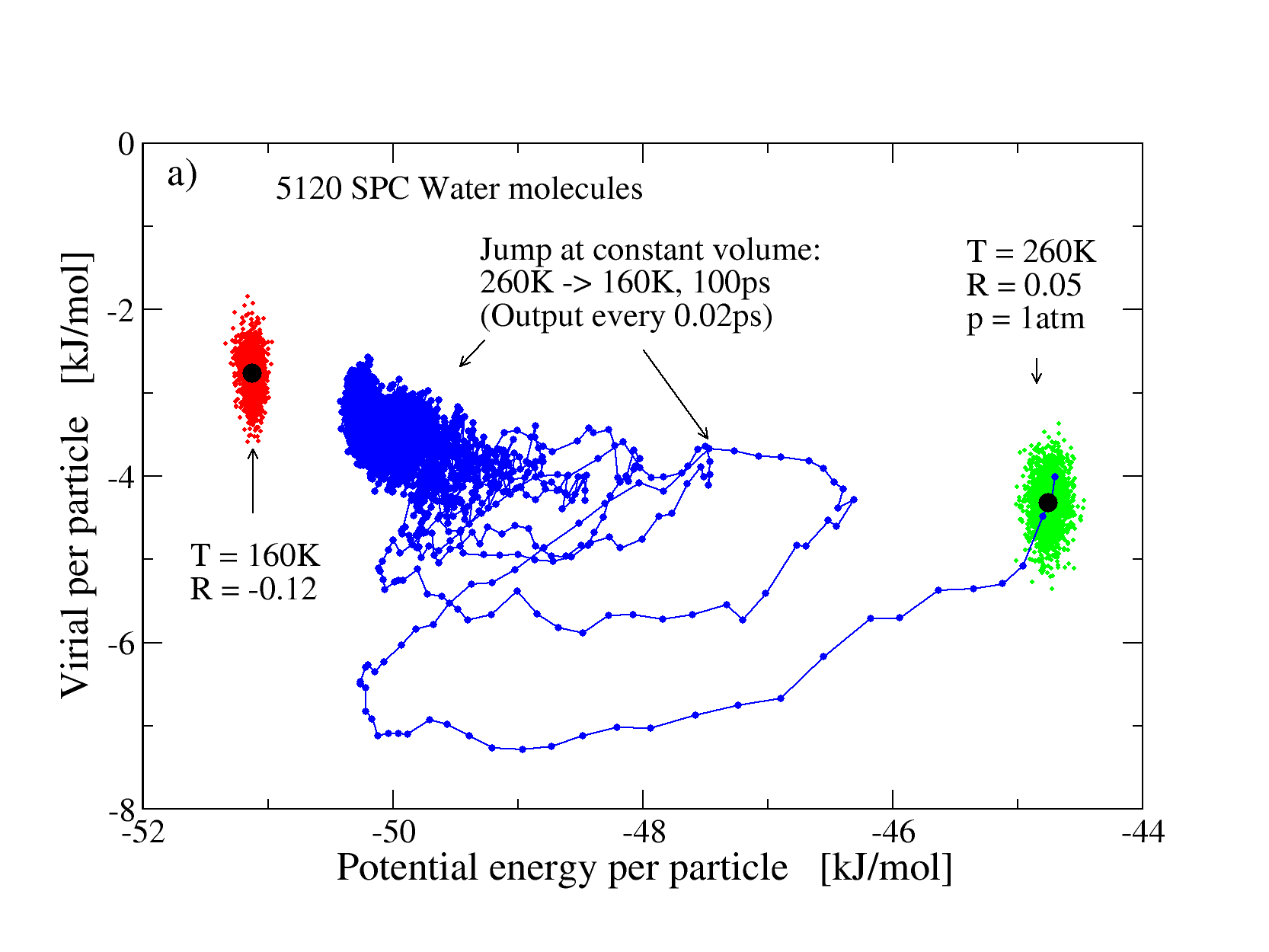}
\includegraphics[width=10cm]{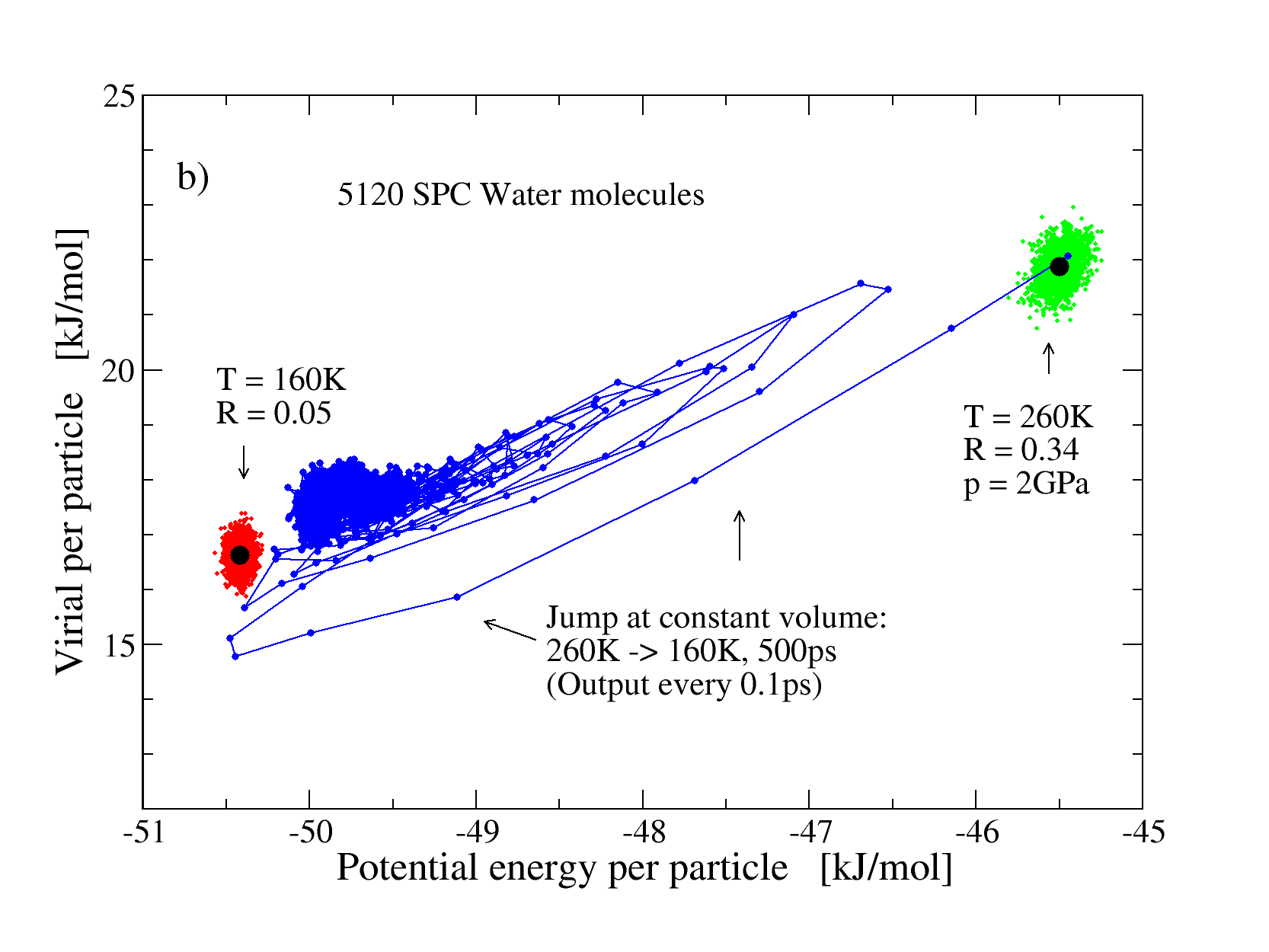}
\end{center}
\caption{(Color) Virial versus potential energy after a temperature down-jump at constant volume applied to SPC water, which is not strongly correlating (colors as in Fig. \ref{WU_aging_scl}). 
(a) SPC water at 1 atm equilibrated at T=260K, subsequently subjected to an isochoric temperature down jump to T=160 K. Clearly $W$ and $U$ are not strongly correlated during the aging process.
(b) Same procedure starting from a 2GPa state point.} 
\label{WU_aging_water}
\end{figure}

According to the extended inverse power-law (eIPL) explanation detailed in the previous section, strong $WU$ correlations characterize all configurations of the LJ liquid at a given volume. This means that the correlations should be there also under non-equilibrium conditions if the volume is kept constant. In this section we present numerical evidence that this prediction is indeed fulfilled, even for molecular models, provided they are strongly correlating in their equilibrium $WU$ correlations.

\subsection{Temperature down-jump simulations of three molecular model liquids}

Figure \ref{WU_aging_scl}(a) shows the results for a temperature down-jump at constant volume, starting and ending in equilibrium (NVT simulations).\footnote{The system consisted of 512 asymmetric dumbbell molecules modelled as two Lennard-Jones spheres connected by a rigid bond. The dumbbells were parameterized to mimic toluene. A large sphere (mimicking a phenyl group) was taken from the Lewis-Wahnstr{\"o}m OTP model \cite{Lewis/Wahnstrom:1994} with the parameters $m_p=77.106$ u, $\sigma_p=0.4963$ nm and $\epsilon_p=5.726$ kJ/mol. A small sphere (mimicking a methyl group) was taken from UA-OPLS having $m_m=15.035$ u, $\sigma_m=0.3910$ nm and $\epsilon_m=0.66944$ kJ/mol. The bonds were kept rigid with a bond length of $d=0.29$ nm. The volume was $V=77.27$ nm$^3$ giving an average pressure of approximately 1 atm. The temperature was held constant at $T=130$ K using the Nos\'{e}-Hoover thermostat. NVT simulations were carried out using Gromacs software.\cite{Berendsen/vanderSpoel/vanDrunen:1995, Lindahl/Hess/vanderSpoel:2001} using the Nos\'{e}-Hoover thermostat.\cite{Nose:1984, Hoover:1985} Molecules were kept rigid using the LINCS algorithm.\cite{Hess/others:1997}} The system studied is an asymmetric dumbbell liquid consisting of two different-sized LJ particles glued together by a bond of fixed length, with parameters chosen to mimic toluene.\cite{Pedersen/others:2008a} The system was first equilibrated at 300 K. The green ellipse consists of several simultaneous instantaneous values of $U$ and $W$ in equilibrium at T=300 K. The strong $WU$ correlation is revealed by the elongation of the ellipse ($R=0.97$; $\gamma = 6.1$). When the liquid is similarly equilibrated at 240 K, the red blob appears. To test for correlation in an out-of-equilibrium situation we changed temperature abruptly from the 300 K equilibrium situation to 240 K. The blue points show how virial and potential energy evolve following the temperature down jump. Clearly, strong $WU$ correlations are present also during the aging towards equilibrium. Figure \ref{WU_aging_scl}(b) shows the same phenomenon for the Lewis-Wahnstr{\"o}m ortho-terphenyl (LW OTP) model which consists of three LJ spheres at fixed length and angle with parameters optimized to mimic ortho-terphenyl.\cite{Lewis/Wahnstrom:1994} This liquid is also strongly correlating ($R=0.91$; $\gamma= 7.6$). The colors are as in Fig. \ref{WU_aging_scl}(a): Green gives a high-temperature equilibrium state ($T=600$ K), red a low-temperature equilibrium state ($T=450$ K), and the blue points show the aging towards equilibrium after changing temperature from 600 K to 450 K. The picture is the same as in Fig. \ref{WU_aging_scl}(a): The blue points follow the dashed line. Thus virial and potential energy correlate strongly also for far-from-equilibrium states. Figure~\ref{WU_aging_scl} (c) plots $W(t)$ and $U(t)$ after the temperature jump for the Fig. \ref{WU_aging_scl}(a) data for the asymmetric dumbbell liquid. $W(t)$ and $U(t)$ follow each other closely on the picosecond time scale as well as in their slow, overall drift to equilibrium. 

What happens when the same simulation scheme is applied to a liquid that is not strongly correlating? An example is SPC water, where the hydrogen bonds are mimicked by Coulomb interactions.\cite{Berendsen/Grigera/Straatsma:1987} Figure~\ref{WU_aging_water} shows results of simulations of SPC water at two different densities, (a) corresponding to low pressure and (b) to very high pressure. In the first case virial and potential energy are virtually uncorrelated ($R=0.05$, $T=260$ K); in the second case correlations are somewhat stronger ($R=0.34$, $T=260$ K), though still weak. As in Fig. \ref{WU_aging_scl} green denotes the initial, high-temperature equilibrium, red the low-temperature equilibrium, and blue the aging towards equilibrium. Clearly, for this system $W$ and $U$ are not closely linked to one another during the relaxation towards  equilibrium.

\subsection{Pressure and energy monitored during crystallization of a supercooled liquid: The Lewis-Wahnstr{\"o}m OTP model}

\begin{figure}
\begin{center}
  \includegraphics[width=8cm]{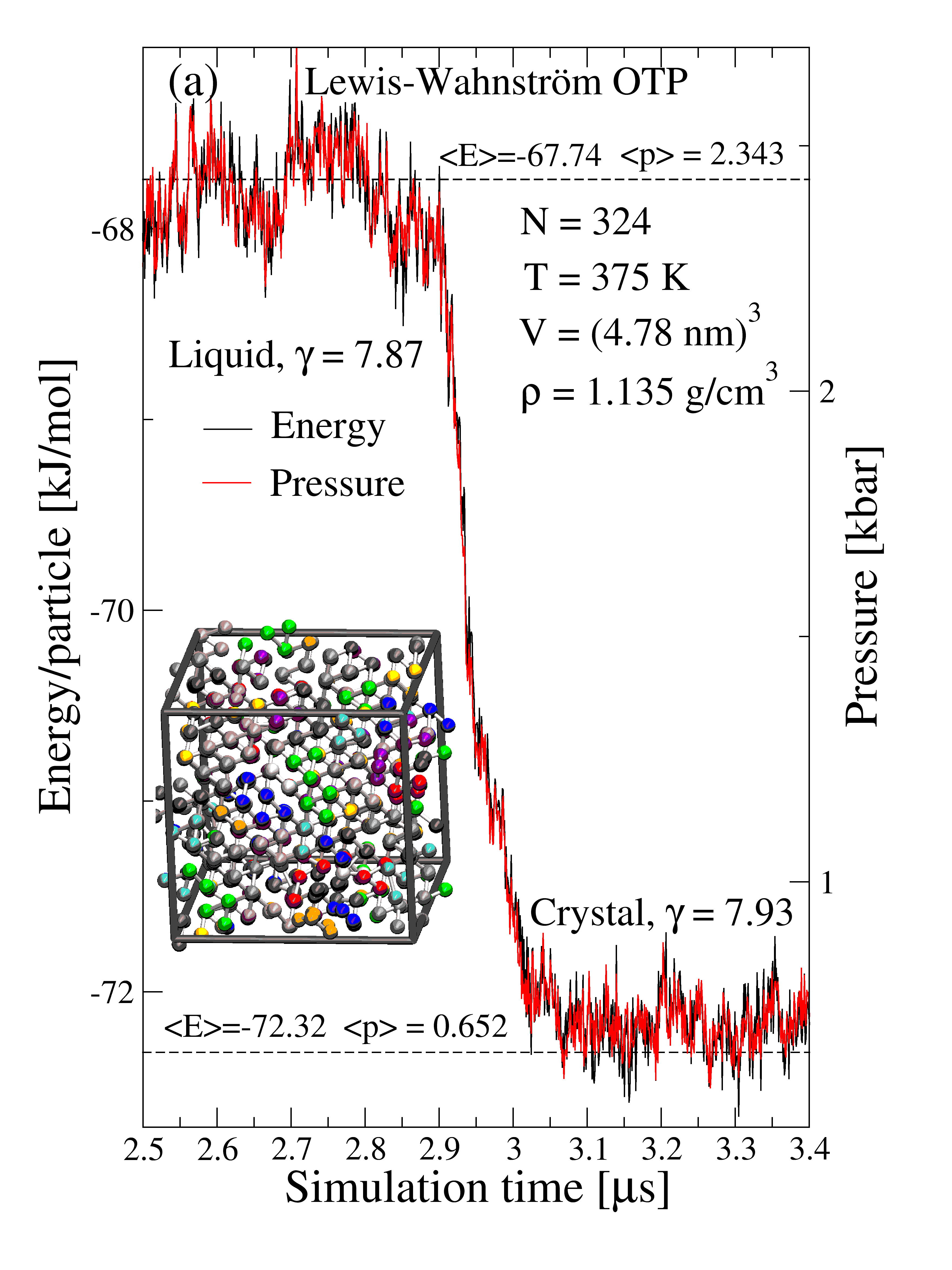}
  \includegraphics[width=8cm]{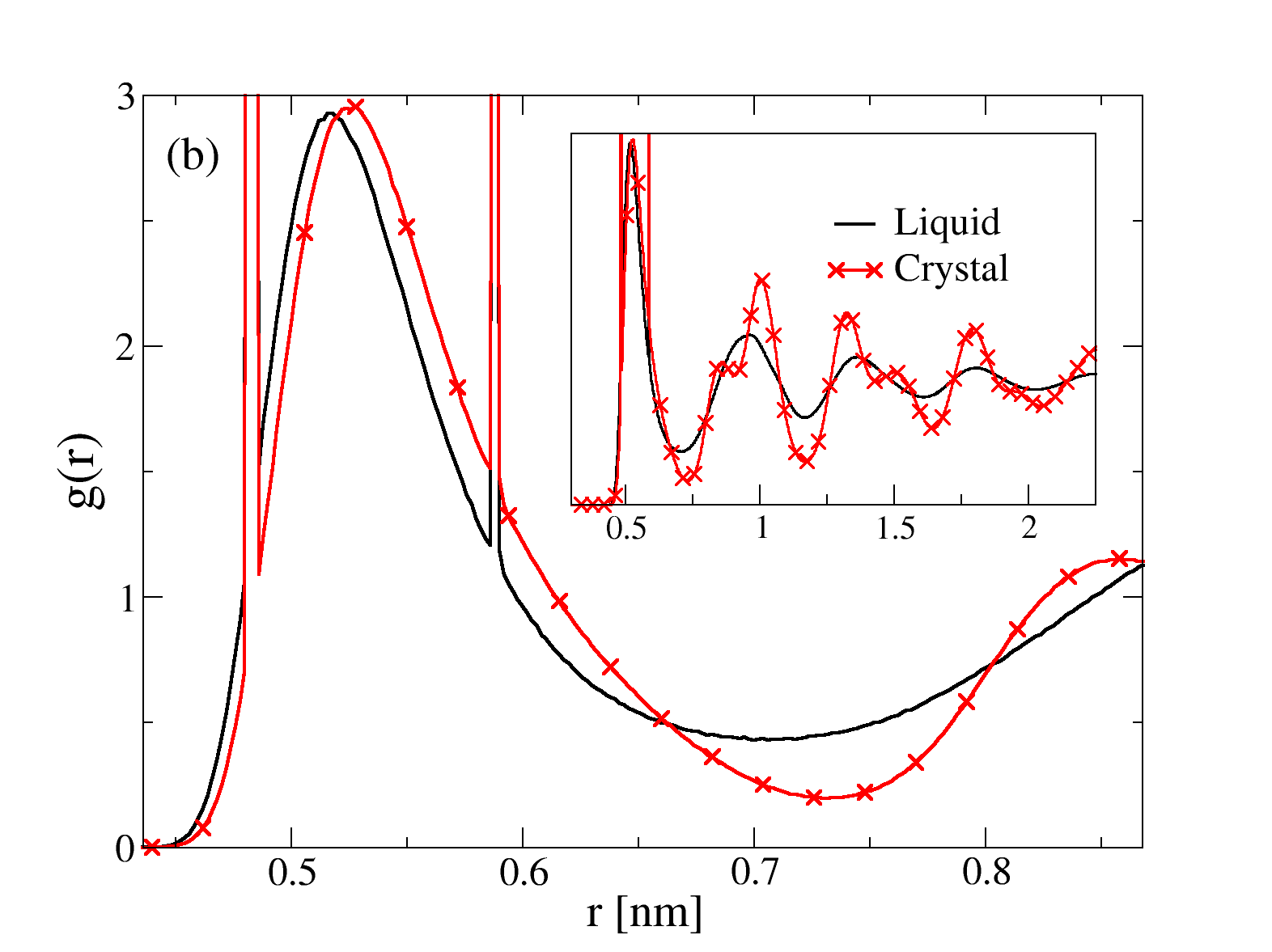}
\end{center}
\caption{(Color) Crystallization of the supercooled Lewis-Wahnstr{\"o}m ortho-terphenyl (OTP) liquid where each molecule consists of three Lennard-Jones spheres with fixed bond lengths and angles.\cite{Lewis/Wahnstrom:1994}
(a) Pressure (right) and energy (left) monitored as functions of time during crystallization at constant volume. Both quantities were averaged over 1 ns; on this time scale the pressure / energy fluctuations directly reflect the virial / potential energy fluctuations. The horizontal dashed lines indicate the liquid (upper line) and crystal (lower line), the averages of which were obtained from the simulation by averaging over times 0-2 $\mu$s and 5-10 $\mu$s, respectively. Both liquid and crystal show strong correlations, and the correlations are also present during crystallization. Inset: Crystal structure from the simulation.
(b) Radial distribution functions of liquid and crystalline phases. The two spikes present in both phases come from the fixed bond lengths.
}
\label{Ulf_figures}
\end{figure}

A different far-out-of-equilibrium situation is that of crystallization of a supercooled liquid monitored at fixed volume and temperature. To the best of our knowledge crystallization of the LW OTP model has not been reported before, but Fig. \ref{Ulf_figures} shows that for simulations over microseconds the supercooled liquid crystallizes at T=375 K and $\rho=1.135 {\rm g/cm^3}$. In the crystal the LJ spheres are arranged in a slightly deformed bcc lattice where the molecules have otherwise random orientation. The crystal is shown in the inset of Fig. \ref{Ulf_figures}(a). Figure \ref{Ulf_figures}(a) shows how time-averaged pressure and energy develop during crystallization. Contributions to pressure and energy from the momentum degrees of freedom are virtually constant after averaging over 1 ns, so strong $WU$ correlations manifest themselves in strong averaged pressure /averaged energy correlations. Clearly averaged pressure and averaged energy follow each other closely also during crystallization. This confirms the above finding, as well as those of Ref. \onlinecite{Bailey/others:2008b}, that strong correlations apply also for the crystalline phase of a strongly correlating liquid. Note that the slope $\gamma$ is virtually unaffected by the crystallization. The persistence of strong virial / potential energy correlation during crystallization -- and the insignificant change of $\gamma$ -- are noticeable, because physical characteristics are rarely unaffected by a first-order phase transition. These simulations show that the property of strong virial / potential energy correlations pertains to the intermolecular potential, not to the particular configurations under study. Figure \ref{Ulf_figures}(b) shows the radial distribution functions for the liquid and crystalline phases.

\subsection{Glasses and inherent states}

\begin{figure}
\begin{center}
  \includegraphics[width=10cm]{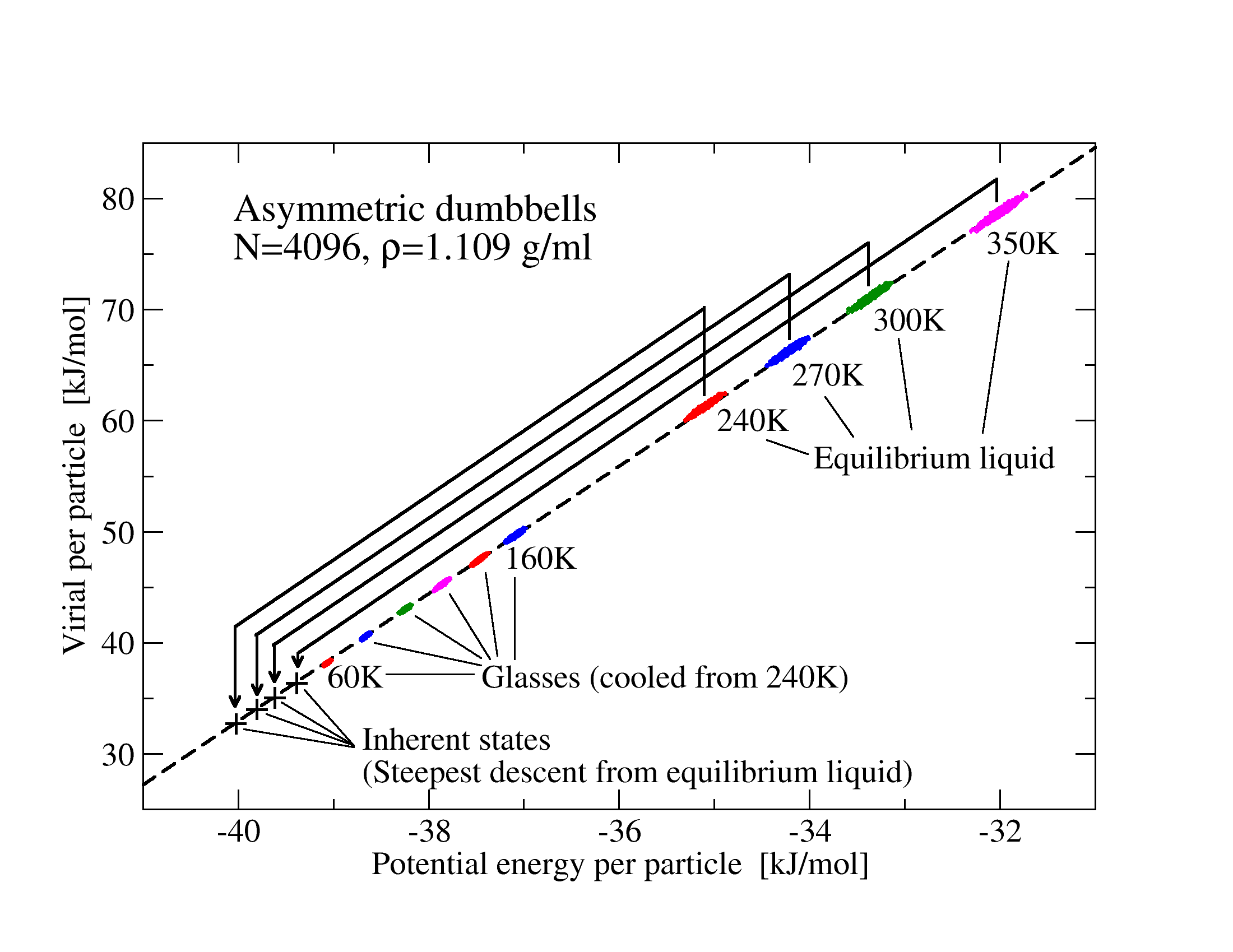}
\end{center}
\caption{(Color) $WU$ plot for the asymmetric dumbbell model for various states at the same volume. The upper right corner shows data for simultaneous values of virial and potential energy for four equilibrium simulations (T=240-350 K). When quenching each of these to zero temperature in order to identify the inherent states, the crosses are arrived at. The intermediate points are glasses prepared by different cooling rates: out-of-equilibrium systems generated by cooling in 1 ns from 240 K to the temperature in question. This plot shows that strong virial / potential energy correlations are not limited to thermal equilibrium situations.
}
\label{WvsEpotLiq_Glass_Inh}
\end{figure}

The above out-of-equilibrium simulations show that the property of strong $WU$ correlation is not confined to thermal equilibrium. This is consistent with the eIPL description given in the previous section, but was shown here to apply even for molecular models with fixed bonds. It appears that strongly correlating liquids have a particularly simple configuration space. These results have significance for any out-of-equilibrium situation. Consider the potential energy landscape picture of viscous liquid dynamics\cite{Goldstein:1969,Stillinger/Weber:1983,Stillinger:1995,Schroder/others:2000} according to which each configuration has an underlying inherent state defined via a deepest-descent quench, a state that contains most information relevant to the slow dynamics, which may be regarded as jumps between different inherent states \cite{Goldstein:1969,Stillinger/Weber:1983,Stillinger:1995,Schroder/others:2000}. Figure \ref{WvsEpotLiq_Glass_Inh} shows a $WU$ plot of the asymmetric dumbbell model in different situations at same density: equilibrium states (upper right) and their corresponding inherent states (lower left, one quench per temperature), and glasses at different temperatures in between. A glass is an out-of-equilibrium state, of course, and inherent states may be regarded as zero-temperature glasses. The plot shows, once again, that strong correlations are present also far from equilibrium.

\section{\label{ENS-DEP}Ensemble dependence of the correlation coefficient}

Most of the simulations of Paper I were done in the NVT ensemble. An obvious question is how the correlations differ between ensembles. This section develops the necessary theory needed to answer this question and compares first the NVT and NVE ensemble, then the NVT and NpT ensembles.

It is well known that although simple thermodynamic averages are independent of ensemble, \footnote{In particular, averages of quantities which are the sums of single-particle  functions.\cite{Allen/Tildesley:1987}} fluctuations are generally ensemble dependent. Consider two ensembles, one with extensive variable $F$ held fixed, and one with its conjugate intensive variable $f$ held fixed (defined so their product is dimensionless). The other parameters defining the ensembles are the same. The covariance of observables $A$ and $B$ in the two ensembles are related as \cite{Allen/Tildesley:1987,Lebowitz/Perkus/Verlet:1967}

\begin{equation}\label{fluc_F_f_relation}
\angleb{\Delta A\Delta B}_F = \angleb{\Delta A\Delta B}_f+
\left(\frac{\partial f}{\partial F}\right)
\left(\frac{\partial}{\partial f}\angleb{A}_f\right)
\left(\frac{\partial}{\partial f}\angleb{B}_f\right)\,.
\end{equation}

\subsection{NVT versus NVE}

To compare the NVT and NVE ensembles we take $F$ and $f$ as the energy $E$ and the inverse temperature $\beta=1/(k_BT)$, respectively, keeping the volume fixed in both cases. Noting that $\partial/\partial\beta=-k_BT^2\partial/\partial T$ and  $(\partial \beta/\partial E)_V = 1/(-k_B T^2 C_V)$ where $C_V$ is the extensive isochoric specific heat, $C_V=Vc_V$, the covariance of $U$ and $W$ is given by 

\begin{align}
\angleb{\Delta U\Delta W}_{NVE} &= \angleb{\Delta U\Delta W}_{NVT}
- \frac{k_BT^2}{C_V}
\frac{\partial\angleb{U}_{NVT}}{\partial T}
\frac{\partial\angleb{W}_{NVT}}{\partial T}
\\
&= \angleb{\Delta U\Delta W}_{NVT} -k_B T^2 C_V^{\textrm{ex}} 
V\beta_V^{\textrm{ex}}/C_V\,.
\end{align}
Here we have introduced the excess parts of the isochoric specific heat and pressure coefficient, $C_V^{\textrm{ex}} $ and $\beta_V^{\textrm{ex}}$, respectively (note the change in notation from Paper II where the superscript ``conf'' was used -- ``ex'' is however more standard in liquid state theory). These two quantities are given by

\begin{align}
C_V^{\textrm{ex}} &= 
\left(\frac{\partial \angleb{U}}{\partial T}\right)_V=
C_V-\frac{3}{2}Nk_B\\
\beta_V^{\textrm{ex}} &= 
\left(\frac{\partial \angleb{W/V}}{\partial T}\right)_V=
\beta_V-\frac{Nk_B}{V}\,.
\end{align}
For the variances one has

\begin{align}
\angleb{(\Delta U)^2}_{NVE} &= \angleb{(\Delta U)^2}_{NVT} -k_B T^2 
(C_V^{\textrm{ex}})^2/C_V \\
\angleb{(\Delta W)^2}_{NVE} &= \angleb{(\Delta W)^2}_{NVT} -k_B T^2 
(V\beta_V^{\textrm{ex}})^2/C_V\,.
\end{align}
The above implies that the NVE $WU$-correlation coefficient $R_{E}$ (in the following subscripts $E$ or $T$ indicate the NVE or NVT ensemble, respectively) is given by

\begin{widetext}

\begin{equation}\label{R_E_fluctuationsT}
R_{E} = \frac{\angleb{\Delta U\Delta W}_{NVT} -k_B T^2 C_V^{\textrm{ex}} 
V\beta_V^{\textrm{ex}}/C_V}
{\sqrt{\angleb{(\Delta U)^2}_{NVT} -k_B T^2 (C_V^{\textrm{ex}})^2/C_V}
\sqrt{\angleb{(\Delta W)^2}_{NVT} -k_B T^2 (V\beta_V^{\textrm{ex}})^2/C_V}}\,.
\end{equation}
We wish to express the right side in terms of the NVT coefficient $R_T$. The definition of $R_T$ implies

\begin{equation}
\angleb{(\Delta W)^2}_{NVT} = \frac{\angleb{\Delta U\Delta W}_{NVT}^2}
{R_T^2 \angleb{(\Delta U)^2}_{NVT}}\,.
\end{equation}
Inserting this into Eq.~(\ref{R_E_fluctuationsT}) and making use of the fluctuation relations $\angleb{\Delta U\Delta W}_{NVT}=k_BT^2V\beta_V^{\textrm{ex}}$ and  $\angleb{(\Delta U)^2}_{NVT}=k_BT^2C_V^{\textrm{ex}}$ (see, e.g., appendix B of Paper I) gives

\begin{equation}
R_{E} = \frac{k_BT^2V\beta_V^{\textrm{ex}} - k_BT^2  C_V^{\textrm{ex}} 
V\beta_V^{\textrm{ex}}/C_V}
{\sqrt{k_BT^2C_V^{\textrm{ex}}-k_B T^2 (C_V^{\textrm{ex}})^2/C_V}
\sqrt{\frac{(k_BT^2V\beta_V^{\textrm{ex}})^2}{R_T^2k_BT^2C_V^{\textrm{ex}}}-k_B T^2 (V\beta_V^{\textrm{ex}})^2/C_V}}\,.
\end{equation}

\end{widetext}
After squaring and cancelling factors of $k_BT^2$ and $V\beta_V^{\textrm{ex}}$ we get an expression relating $R_E$ to $R_T$,

\begin{equation}\label{R_E_R_T}
R_E^2 = 
\frac{ 1-C_V^{\textrm{ex}}/C_V} 
{1/R_T^2 - C_V^{\textrm{ex}}/C_V}\,.
\end{equation}
To get a feel for the relation, divide by $R_T^2$. This yields

\begin{equation}
\left(\frac{R_E}{R_T}\right)^2 
= \frac{1 - C_V^{\textrm{ex}} / C_V}{1 - R_T^2 C_V^{\textrm{ex}} / C_V}.
\end{equation}
First one notes that when $R_T=1$, the denominator on the right side becomes equal to the numerator and $R_E=1$. That is, the property of perfect correlation is independent of (fixed-volume) ensemble. When $R_T< 1$, the denominator becomes greater than the numerator, and so $R_E^2<R_T^2$. That is, the correlation coefficient is smaller in the NVE ensemble than in the NVT one. How can we understand this? Consider the set of $WU$ points sampled by the system during an NVE trajectory; this is an elongated blob in the $WU$ diagram. Changing the energy will cause the blob to move along a line almost parallel with the long axis of the blob.\cite{Bailey/others:2008b} Switching to the NVT ensemble is equivalent to superposing several of these collinear blobs on top of each other---the result is necessarily longer, but not wider. This corresponds to higher correlation.

Simulation data confirming relation (\ref{R_E_R_T}) are presented in Table~\ref{compareRERT}.

\begin{table}
\begin{tabular}{|c|c|c|c|c|c|c|}
\hline
system & $\rho$ & $T$ & $R_{T}$ & $C_V^{\textrm{ex}}/N$ & Eq.~(\ref{R_E_R_T})
 & $R_{E}$ \\
\hline
LJ     &  1.00 & 1.00 & 0.991  & 1.5  & 0.982 & 0.983 \\ 
LJ     &  1.00 & 0.80 & 0.991  & 1.7  & 0.981 & 0.981 \\
LJ     &  0.82 & 0.80 & 0.943  & 0.90 & 0.912 & 0.918 \\
LJ     &  0.82 & 0.67 & 0.949  & 1.3 & 0.909  & 0.904 \\
KABLJ  &  1.2  & 0.47 & 0.936  & 2.1 & 0.859 & 0.862 \\
\hline
\end{tabular}
\caption{\label{compareRERT}Check of relation (\ref{R_E_R_T}) between $R_T$ and $R_E$ for the LJ and the Kob-Andersen binary Lennard-Jones (KABLJ) fluids. The units for $\rho$ and $T$ are the dimensionless units defined in terms of the length and energy parameters $\sigma$ and $\epsilon$ for the interactions of the large particles. The excess isochoric heat capacity was calculated from the potential energy fluctuations in NVT ensemble.}
\end{table}

\subsection{NVT versus NpT}

In the NpT ensemble, where volume is allowed to fluctuate, we must consider different variables. The natural variables to correlate are the excess enthalpy $H_{\rm ex}\equiv U+pV$ and the volume $V$. We use again Eq.~(\ref{fluc_F_f_relation}), but now take $F$ as $V$ and $f$ as $p\beta$, the pressure times inverse temperature, keeping temperature constant. Eq.~(\ref{fluc_F_f_relation}) becomes

\begin{equation}\label{fluc_V_p_relation}
\angleb{\Delta A\Delta B}_{NVT} = \angleb{\Delta A\Delta B}_{NpT}-
\frac{k_BTK_T}{V}
\left(\frac{\partial}{\partial p}\angleb{A}\right)_T
\left(\frac{\partial}{\partial p}\angleb{B}\right)_T.
\end{equation}
The details of the calculation, which are somewhat tedious, are given in Appendix A. The result for the $H_{\rm ex}V$ correlation coefficient is rather simple, though:

\begin{equation}
R_{H_{\rm ex}V,NpT} = \frac{1}{\sqrt{1+b^2/a^2}},
\end{equation}
where

\begin{equation}
a=\angleb{\Delta U\Delta W}_{NVT} + N(k_BT)^2,
\end{equation}
and

\begin{equation}
b^2=K_TVk_BT\angleb{(\Delta U)^2}_{NVT}.
\end{equation}
Notice that $R_{H_{\rm ex}V,NpT}$ is strictly less than unity -- even for perfectly correlating liquids (that is, with perfect $WU$ correlations in the NVT and NVE ensembles). For the Lennard-Jones simulation of Fig.~\ref{WU_NVT_NVP} (b)  $R_{H_{\rm ex}V,NpT}=0.86$ (recall the NVT $W,U$ correlation coefficient for the same state point is 0.94). Unlike the situation when comparing the NVT and NVE ensembles, there does not seem to be a simple relation between the two correlation coefficients. It seems likely, though, that the $H_{\rm ex}V$ correlation in the NpT ensemble is generally smaller than the $WU$ correlation in the NVT ensemble; thus the property of strong correlation is less evident in the NpT ensemble.

\section{\label{THERMO}Thermodynamics of strongly correlating liquids}

The property of strong virial / potential energy correlation not just refers to microscopic properties that are only accessible in simulation, it also has consequences for the liquid's thermodynamics as well. The first subsection below relates the slope (Eq. (\ref{slopeDefinition})) to the Gr{\"u}neisen parameter, the second subsection shows how to give a general thermodynamic formulation of the property of strong correlations.

\subsection{\label{Gruneisen}Relation to the Gr\"uneisen parameter}

The Gr\"uneisen parameter was originally introduced to characterize the volume dependence of normal modes of a crystal:\cite{Born/Huang:1954,Ashcroft/Mermin:1976}

\begin{equation}
\gamma_i = -\frac{d\ln(\omega_i)}{d\ln(V)}\,,
\end{equation}
where $\omega_i$ is the frequency of the $i$th normal mode. By assuming that $\gamma_i$ is the same for all modes and denoting the common value by $\gamma_G$, one can derive the Mie-Gr\"uneisen equation of state,\cite{Born/Huang:1954} 

\begin{equation}\label{MieGruneisen}
p + \frac{du}{dv} = \gamma_G \frac{E_{\rm vib}}{V}\,.
\end{equation}
Here $p$ is pressure, $u(v)$ with $v=V/N$ is the ``static'' energy of the crystal per atom (the energy of the force-free configuration about which vibrational motion occurs), and $E_{\rm vib}$ is the vibrational energy. In general $\gamma_G$ depends on volume, but this dependence is typically small enough that it can be neglected. From Eq.~(\ref{MieGruneisen}) it follows that, if $E$ is the total, thermally averaged internal energy, one has 
$(\partial p / \partial T)_V=(\gamma_G/V)(\partial E_{\rm vib} / \partial T)_V=(\gamma_G/V)(\partial E / \partial T)_V$, i.e., 

\begin{equation}\label{gammaG_pressureDeriv}
\gamma_G = V\left(\frac{\partial p}{\partial E}\right)_V\,.
\end{equation}
This expression is the slope of the pressure versus energy curve at fixed volume, analogous to the $\gamma$ of Eq. (\ref{slopeDefinition}) but for the presence of the kinetic terms (recall that for an IPL liquid $\gamma=n/3=(\partial W/\partial U)_V$). If $\alpha_p$ is the coefficient of thermal expansion, $K_T$ the isothermal bulk modulus, and $c_V=C_V/V$ the isochoric specific heat per unit volume, Eq. (\ref{gammaG_pressureDeriv}) implies via standard thermodynamic identities

\begin{equation}\label{thermodynamicGrunseisen}
\gamma_G = \frac{\alpha_p K_T}{c_V}\,.
\end{equation}
This relation allows $\gamma_G$ to be determined from experimentally accessible quantities; in fact Eq.~(\ref{thermodynamicGrunseisen}) can be taken as a thermodynamic definition of $\gamma_G$.\cite{Wallace:1972}

There have been suggestions of how to connect the so-called density scaling exponent\cite{Alba-Simionesco/others:2004,Roland/others:2005} -- the one controlling the relaxation time via the variable $\rho^{\gamma}/T$ -- with the Gr\"uneisen parameter, notably by Roland and coworkers.\cite{Casalini/Mohanty/Roland:2006,Roland/Feldman/Casalini:2006,Roland/Casalini:2007} In Ref.~\onlinecite{Casalini/Mohanty/Roland:2006} equality of $\gamma_G$ and $\gamma$ was argued theoretically. More recently, Roland and Casalini  showed\cite{Roland/Casalini:2007} that equality is not consistent with experimental results; rather $\gamma_G$ is smaller than $\gamma$ by a factor of order three. This discrepancy was reconciled in the context of the entropy model for relaxation (by which the relaxation time is a unique function of the so-called configurational entropy $S_c$), by introducing a corrected Gr\"uneisen parameter defined via

\begin{equation}
\gamma_G^{\textrm{corr}} = \frac{\alpha_p K_T}{\Delta c_V}\,.
\end{equation}
Here $\Delta c_V$ is the difference of the isochoric specific heats per unit volume between the liquid and the glass. Because $\Delta c_V$ is smaller than $c_V$, one has $\gamma_G^{\textrm{corr}}> \gamma_G$. By arguing from experimental data that the non-configurational part of the entropy (associated with vibrations, equal to the entropy of the glass) is independent of volume and assuming that $\Delta C_V$ is constant, they derive 

\begin{equation}
S_c = \Delta C_V\ln(TV^{\gamma_G^{\textrm{corr}}}) + \textrm{const}\,.
\end{equation}
Hence $\gamma_G^{\textrm{corr}}$ is the density-scaling exponent.

We take a different approach to connecting the Gr\"uneisen parameter with the slope $\gamma$ (which provides a good estimate of the density scaling exponent, see Ref. \onlinecite{Schroder/Pedersen/Dyre:2008a}). Instead of splitting the entropy, we split the pressure, into potential and kinetic parts and get from $\gamma_G =V(\partial p/\partial E)_V$ 

\begin{equation}
\gamma_G 
= V\frac{\left(\frac{\partial p}{\partial T}\right)_V}{\left(\frac{\partial E}{\partial T}\right)_V}
=\frac{\left(\frac{\partial W}{\partial T}\right)_V + Nk_B}{\left(\frac{\partial E}{\partial T}\right)_V}\,.
\end{equation}
Expressing the temperature derivatives in terms of fluctuations (Appendix B of Paper I) gives

\begin{equation}
\gamma_G = \frac
{ \angleb{\Delta U\Delta W}/k_BT^2 + Nk_B}
{ C_V}\,.
\end{equation}
In the limit of strong correlation one can replace $\angleb{\Delta U\Delta W}$ with $\gamma\angleb{(\Delta U)^2}$. Writing the resulting expression in terms of the excess (configurational) specific heat 
$C_V^{\textrm{ex}}$ gives

\begin{equation}
\gamma_G = \frac{ \gamma C_V^{\textrm{ex}} + Nk_B}
{ C_V}\,.
\end{equation}
In the harmonic approximation, good for many simple liquids close to their melting point,\cite{Chisolm/Wallace:2001} $C_V^{\textrm{ex}}=(3/2)Nk_B$ (while it is generally larger in the supercooled liquid state). Thus the term $Nk_B$ in the numerator is expected to be roughly a factor of ten smaller than the other term; we drop it and arrive at

\begin{equation}
\frac{\gamma_G}{\gamma} \cong  \frac{C_V^{\textrm{ex}}}{C_V}\,.
\end{equation}
This  ratio is around one half in the harmonic approximation, otherwise larger.

\subsection{\label{EBF}Energy-bond formulation of the strongly correlation property}

This section derives a general thermodynamic condition of the property of strong $WU$ correlations, a condition which linearly constrains small variations in entropy, volume, temperature and pressure (Eq. (\ref{scl_constraint}) below). It is convenient to approach the problem from a general point of view. The energy-bond formalism provides an abstract description of the interactions between a system and its surroundings.\cite{Paynter:1961,Oster/Perelson/Katchalsky:1971,Oster/Perelson/Katchalsky:1973,Christiansen:1978,Christiansen:1979,Mikulecky:1993,Karnopp/Margolis/Rosenberg:2006} An energy bond has an ``effort'' variable $e(t)$ and a ``flow'' variable $f(t)$, where $e(t)f(t)$ is the free energy transferred into the system per unit time. The ``displacement'' $q(t)$ is the time-integrated flow, i.e., $\dot q(t)=f(t)$. The energy-bond formalism is general, but we only discuss the linear case where it is most useful. Thus we consider a system that is slightly perturbed from equilibrium. It is assumed that the underlying microscopic dynamics is described by a stochastic equation, i.e., inertial forces are ignored.

Linear-response theory is characterized by the fluctuation-dissipation (FD) theorem, which in the energy-bond formalism is given as follows. Consider a situation with $n$ energy bonds and external control of the effort variables. If $\langle f_i(0) f_j(t')\rangle_0$ is the equilibrium flow autocorrelation function, the average flow at time $t$ is given by

\begin{equation}\label{fd_e}
f_i(t)\,=\,\frac{1}{k_BT}\sum_{j=1}^n\,
\int_0^\infty \langle f_i(0) f_j(t')\rangle_0 e_j(t-t')dt'\,.
\end{equation}
If the arbitrary additive constants of the displacements are chosen such that $\la q_i\ra_0=0$, the time-integrated version of this is

\begin{equation}\label{fd_q}
q_i(t)\,=\,-\frac{1}{k_BT}\sum_{j=1}^n\, 
\int_0^\infty \langle q_i(0) f_j(t')\rangle_0 e_j(t-t')dt'\,.
\end{equation}
If the flow variables are externally controlled, the FD theorem is 

\begin{equation}\label{fd_f}
e_i(t)\,=\,\frac{1}{k_BT}\sum_{j=1}^n\,
\int_0^\infty \langle e_i(0) e_j(t')\rangle_0 f_j(t-t')dt'\,.
\end{equation}
In most cases efforts are invariant under time reversal and flows change sign. The Onsager reciprocity relation is $\langle f_i(0) f_j(t)\rangle_0=\langle f_j(0) f_i(t)\rangle_0$ (or $\langle e_i(0) e_j(t)\rangle_0=\langle e_j(0) e_i(\tau)\rangle_0$, depending on which variables are externally controlled and which are free to fluctuate). From the FD theorem expressions for the frequency-dependent response functions are easily derived. Consider for instance the compliances $J_{ij}(\omega)$, defined by $J_{ij}(\omega)=\delta q_i(\omega)/\delta e_j(\omega)$ for a periodic situation with linear perturbations around equilibrium, $e(t)={\rm Re}[e(\omega)\exp(i\omega t)]$, etc. For these quantities the FD theorem implies 

\begin{equation}\label{fd}
J_{ij}(\omega)\,=\,-\frac{1}{k_BT}\,\int_0^\infty \langle q_i(0) f_j(t')\rangle_0 \exp(-i\omega t')dt'\,.
\end{equation}

The case relevant to strongly correlating liquids is that of two energy bonds which are not independent, as we now proceed to show. The two energy bonds are those of standard thermodynamics, reflecting the fundamental relation $dE=TdS-pdV$:\cite{Ellegaard/others:2007,Christensen/Dyre:2008} The thermal energy bond with temperature variation as the effort variable and entropy variation as the displacement variable ($e_1(t)=\delta T(t)$, $q_1(t)=\delta S(t)$), and the mechanical energy bond with pressure variation as the effort variable and the negative volume variation as the displacement variable ($e_2(t)=\delta p(t)$, $q_2(t)=-\delta V(t)$). Usually the two standard thermodynamic energy bonds are independent, but we are here interested in the case when they are not.

Treating the problem of two constrained energy bonds from a general perspective, we shall prove that the following four criteria are equivalent:

\begin{enumerate}
\item The variables of the two energy bonds  are linearly constrained as follows

\begin{equation}\label{constraint}
a\,q_1(t)+b\,q_2(t)\,=\,
c\,e_1(t)+d\,e_2(t)\,.
\end{equation}

\item
The system's relaxing properties, i.e., its non-instantaneous responses, are described\cite{Ellegaard/others:2007} by a single variable $\epsilon(t)$ as follows:

\begin{eqnarray}\label{sop_e}
q_1(t)\,&=&\,
J_{11}^\infty e_1(t)+J_{12}^\infty e_2(t)+\gamma_1\epsilon(t)\nonumber\\
q_2(t)\,&=&\,
J_{21}^\infty e_1(t)+J_{22}^\infty e_2(t)+\gamma_2\epsilon(t)\,.
\end{eqnarray}
In these equations the $J^\infty $'s are the compliances referring to the short-time, non-relaxing response (the high-frequency response). Note that $J_{12}^\infty =J_{21}^\infty $ by the FD theorem.

\item
The relaxing parts of the three correlation functions entering into Eq. (\ref{fd}) are proportional. More precisely, the correlation functions obey

\begin{equation}\label{f_autcorr}
\la q_1(0)f_1(t)\ra_0
\,\propto\,
\la q_1(0)f_2(t)\ra_0
\,\propto\,
\la q_2(0)f_1(t)\ra_0
\,\propto\,
\la q_2(0)f_2(t)\ra_0\,\,\, (t \neq 0)\,,
\end{equation}
and

\begin{equation}\label{q_autcorr}
\la q_1(0)f_1(t)\ra_0\la q_2(0)f_2(t)\ra_0
\,=\,
\la q_1(0)f_2(t)\ra_0  \la q_2(0)f_1(t)\ra_0\,\,\, (t \neq 0)\,.
\end{equation}
Note that by differentiation Eq. (\ref{f_autcorr}) implies for $t \neq 0$ that
$\la f_1(0)f_1(t)\ra_0\propto\la f_1(0)f_2(t)\ra_0\propto\la f_2(0)f_1(t)\ra_0\propto\la f_2(0)f_2(t)\ra_0$.

\item
The dynamic Prigogine-Defay ratio\cite{Ellegaard/others:2007} $\Lambda(\omega)$ is unity at all frequencies (where double prime denotes the negative imaginary part):

\begin{equation}\label{pdf}
\Lambda(\omega)\,\equiv\,
\frac{J_{11}''(\omega)J_{22}''(\omega)}{\left(J_{12}''(\omega)\right)^2}\,=\,1\,.
\end{equation}

\end{enumerate}

{\it Proof that $1 \Leftrightarrow 2$}: 
By elimination of the variable $\epsilon$ from Eq. (\ref{sop_e}), $2$ implies $1$. To prove the reverse implication, suppose that Eq. (\ref{constraint}) applies and fix the dimensions such that the constants $c$ and $d$ are dimensionless. Define $J_{11}^\infty = (1+c)/a$, $J_{12}^\infty = J_{21}^\infty = -1/b$, and $J_{22}^\infty = (db+a)/b^2$. Introducing the variables $\epsilon_1=q_1-J_{11}^\infty e_1-J_{12}^\infty e_2$ and $\epsilon_2=q_2-J_{21}^\infty e_1-J_{22}^\infty e_2$, it follows that $a\epsilon_1+b\epsilon_2=0$. This means that we are in the situation described by Eq. (\ref{sop_e}) with a common relaxing variable to the two energy bonds, $\epsilon(t)\propto\epsilon_1(t)\propto\epsilon_2(t)$, and symmetric short-time compliances, $J_{12}^\infty = J_{21}^\infty$.

{\it Proof that $2 \Rightarrow 3$}: In terms of functional derivatives with respect to the efforts at an earlier time $(t'< t)$, since $\epsilon(t)$ for small variations in the effort variables is linear in these, via the FD theorem time-reversal invariance implies that $\delta q_1(t)/\delta e_2(t')=\delta q_2(t)/\delta e_1(t')$. Thus Eq. (\ref{sop_e}) implies $\delta\epsilon(t)/\delta e_2(t') \propto \delta\epsilon(t)/\delta e_1(t')$. From this Eqs. (\ref{f_autcorr}) and (\ref{q_autcorr}) now follow via the FD theorem.

{\it Proof that $3 \Rightarrow 4$}: According to the FD theorem the compliance matrix imaginary parts are given by $J_{ij}''(\omega)=(1/k_BT)\int_0^\infty \langle q_i(0) f_j(t')\rangle_0 \sin(\omega t')dt'$. In conjunction with Eqs. (\ref{f_autcorr}) and (\ref{q_autcorr}) this implies that $\Lambda(\omega)=1$ at all frequencies.

{\it Proof that $4 \Rightarrow 2$}: We refer to the calculations of Ref. \onlinecite{Ellegaard/others:2007} which considered a system described by stochastic dynamics, i.e., with no inertial forces. Generalization of the arguments given there for the two standard thermodynamic energy bonds to the case of two arbitrary energy bonds proves the required implication.

This completes the proof of the equivalence of points 1-4. -- For the case where the two energy bonds are the fundamental thermodynamic bonds, the constraint Eq. (\ref{constraint}) translates into (changing here the sign of $b$)

\begin{equation}\label{scl_constraint}
a\,\delta S(t)+b\,\delta V(t)\,=\,
c\,\delta T(t)+d\,\delta p(t)\,.
\end{equation}

How does this all relate to strong $WU$ correlations in liquids? Via the equivalence of Eq. (\ref{scl_constraint}) to Eq. (\ref{sop_e}) and to unity dynamic Prigogine-Defay ratio (Eq. (\ref{pdf})), the results derived in Refs. \onlinecite{Pedersen/others:2008,Pedersen/others:2008a,Bailey/others:2008a} imply that Eq. (\ref{scl_constraint}) describes a 100\% correlating liquid subjected to small perturbations from equilibrium. Generally, for any strongly correlating liquid Eq. (\ref{scl_constraint}) is obeyed with good accuracy. Thus Eq. (\ref{scl_constraint}) gives the required thermodynamic formulation of the property of the hidden scale invariance characterizing strongly correlating liquids. 

Equation (\ref{scl_constraint}) implies that for strongly correlating liquids the four thermodynamic variables, entropy, volume, temperature, and pressure cannot vary independently. Referring to Eq. (\ref{sop_e}), it is clear that for certain simultaneous changes of the four thermodynamic variables, the relaxing part is left unchanged; this suggests that for such changes the system is taken to a state where it is immediately in thermal equilibrium. This observation inspired the works leading to Paper IV where ``isomorphs'' are introduced. These are curves in the state diagram along which several quantities are invariant, and along which jumps from equilibrium at one state point take the system to a new state that is instantaneously in thermal equilibrium.

Finally we would like to draw attention to an analogue of strongly correlating liquids. Consider a relaxing dielectric such as, e.g., a highly viscous dipolar liquid placed in a metal capacitor. This system's interaction with its surroundings may be described by two energy bonds: One energy bond is defined by the capacitor charge (electronic plus induced) and the voltage across the capacitor, the other energy bond is the induced dielectric charge at the capacitor surface and a fictive electric field only coupling to the liquid's dipoles. Because of Gauss' law these two energy bonds are not independent, but constrained by a linear displacement-field relation of the form Eq. (\ref{constraint}). Thus from the energy-bond formalism point of view, a strongly correlating liquid is analogous to the standard measuring cell used for probing $\epsilon(\omega)$ of dipolar viscous liquids, with the strong virial / potential energy correlations reflecting one of Maxwell's four equations.

\section{Concluding remarks}

We have illuminated a number of features of strongly correlating liquids' hidden scale invariance. The linear term in the eIPL potential, which hides this approximate scale invariance, contributes little to the thermal fluctuations at fixed volume; this is why strongly correlating liquids inherit a number of IPL properties. As shown in previous papers\cite{Pedersen/others:2008,Pedersen/others:2008a,Bailey/others:2008a, Bailey/others:2008b,Bailey/others:2008c} the hidden scale invariance has important experimental consequences, including that of density scaling.\cite{Schroder/Pedersen/Dyre:2008a,cos09} The general physical picture that we would like to suggest is that van der Waals liquids and most or all metallic liquids -- because they are strongly correlating -- are simpler than hydrogen-bonding liquids, ionic liquids, and covalently bonded liquids, which are not strongly correlating. 

Paper IV further investigates the consequences of a liquid being strongly correlating. This is done by defining ``isomorphs'' in the liquid's state diagram and showing that a number of properties to a good approximation are invariant along isomorphs. The isomorph definition does not refer to $WU$ correlations. Only strongly correlating liquids have isomorphs, however; this is because the existence of isomorphs is a direct consequence of the hidden scale invariance of strongly correlating liquids.

\acknowledgments

We thank Tage Christensen and S{\o}ren Toxv{\ae}rd for helpful input. The centre for viscous liquid dynamics ``Glass and Time'' is sponsored by the Danish National Research Foundation (DNRF).

\appendix

\section{Calculating the $H_{\rm ex},V$ correlation coefficient in the NpT ensemble}

Here we provide the details of the calculation of the correlation coefficient between volume and excess enthalpy in the NpT ensemble. We apply Eq.~(\ref{fluc_F_f_relation}) with $\{A,B\} \in \{U,V\}$. First we need the pressure derivatives at constant temperature of $\angleb{U}$ and $\angleb{V}$. Taking $U$ first, we have (noting that for simple averages like $\angleb{U}$, it is not necessary to specify the ensemble because of equivalence of ensembles)

\begin{align}
\left.\frac{\partial \angleb{U}}{\partial p}\right|_T &= 
\left.\frac{\partial \angleb{V}}{\partial p}\right|_T
\left.\frac{\partial \angleb{U}}{\partial V}\right|_T \\
&= -\frac{\angleb{V}}{K_T} 
\left.\frac{\partial \angleb{U}}{\partial V}\right|_T\,.
\end{align}
Note that $V$ in the derivative is without averaging signs since there it is a parameter of the relevant ensemble (NVT). The volume derivative of $\angleb{U}$ is calculated as follows. The excess (configurational) partition function $Z(V,T)$ is the integral

\begin{equation}
Z(V,T) = \int_\Gamma \exp(-\beta U(\Gamma,V)) d\Gamma.
\end{equation}
Here $\Gamma$ indexes points in configuration space and $d\Gamma=d^{3N}r/V^N$. In the following we use the configuration space identity $\partial U/\partial V=-W/V$; constant temperature is implicit, as is the dependence of $U$ on $\Gamma$ and $V$,

\begin{align}
\frac{\partial \angleb{U}}{\partial V} &= 
\frac{\partial}{\partial V}\left( Z^{-1} \int_\Gamma U
\exp(-\beta U)  d\Gamma \right) \\
&=  Z^{-1} \int_\Gamma \left( \frac{\partial U}{\partial V}
+ U(-\beta) \frac{\partial U}{\partial V} \right) 
\exp(-\beta U) d\Gamma  \nonumber \\
& -\frac{1}{Z^2} \left(\int_\Gamma U \exp(-\beta U) 
 d\Gamma \right) \frac{\partial Z}{\partial V} \\
&= \angleb{\frac{\partial U}{\partial V}} -\beta \angleb{U 
\frac{\partial U}{\partial V}} - \angleb{U} Z^{-1}
 \int_\Gamma(-\beta) \frac{\partial U}{\partial V}
\exp(-\beta U) d\Gamma \\
&= \angleb{\frac{\partial U}{\partial V}}
-\beta \angleb{U \frac{\partial U}{\partial V}}
 +\beta \angleb{U}\angleb{\frac{\partial U}{\partial V}} \\
&=  \angleb{\frac{\partial U}{\partial V}} - \beta
\angleb{\Delta U\Delta \left(\frac{\partial U}{\partial V}\right)} \\
&= \frac{1}{V}\left( -\angleb{W} 
+ \beta \angleb{\Delta U\Delta W} \right)\,.
\end{align}
Thus we have (adding the subscript $NVT$ to the fluctuation expression since this is ensemble-dependent)

\begin{equation}
\left.\frac{\partial \angleb{U}}{\partial p}\right|_T = 
\frac{\angleb{W}}{K_T} - 
\frac{\angleb{\Delta U\Delta W}_{NVT}}{k_BTK_T}.
\end{equation}
The pressure dependence of $\angleb{V}$ is given by

\begin{equation}
\left.\frac{\partial\angleb{V}}{\partial p}\right|_T = -\frac{\angleb{V}}{K_T}.
\end{equation}
To keep the notation simple, averaging signs are henceforth omitted from simple averages such as $\angleb{V}$, $\angleb{W}$, etc. We can write expressions for the variances of $U$ and $V$ in the NpT
 ensemble using Eq.~(\ref{fluc_V_p_relation})

\begin{align}
\angleb{(\Delta U)^2}_{NpT}  &= \angleb{(\Delta U)^2}_{NVT} + \frac{k_BTK_T}{V}
\left( \frac{W}{K_T} - \frac{\angleb{\Delta U\Delta W}_{NVT}}{k_BTK_T}\right)^2 \\
\angleb{(\Delta V)^2}_{NpT} &= 0 + \frac{k_BTK_T}{V} \frac{V^2}{K_T^2} = 
\frac{Vk_BT}{K_T}.
\end{align}
We need also the covariance

\begin{align}
 \angleb{\Delta U\Delta V}_{NpT} &= 0 + \frac{k_BTK_T}{V}
\left(\frac{W}{K_T} - \frac{\angleb{\Delta U\Delta W}_{NVT}}
{k_BTK_T}\right)\frac{-V}{K_T} \\
&= -\frac{k_BTW}{K_T} +
\frac{\angleb{\Delta U\Delta W}_{NVT}}{K_T} \label{DeltaUV_NpT}\,.
\end{align}

Now we have all we need to construct the $H_{\rm ex}V$ correlation coefficient in the NpT ensemble. The covariance between $H_{\rm ex}$ and $V$ is

\begin{align}
\angleb{\Delta H_{\rm ex}\Delta V}_{NpT} &= \angleb{\Delta U\Delta V}_{NpT} 
+ p\angleb{(\Delta V)^2}_{NpT} \\
&= \frac{\angleb{\Delta U\Delta W}_{NVT}}{K_T} - \frac{k_BTW}{K_T} 
+ p\frac{Vk_BT}{K_T} \\
&= \frac{\angleb{\Delta U\Delta W}_{NVT}}{K_T} + \frac{k_BT (Nk_BT)}{K_T}\,,
\end{align}
where we have used $pV=Nk_BT+W$. The variance of $H_{\rm ex}$ is more tedious:

\begin{align}
\angleb{(\Delta H_{\rm ex})^2}_{NpT} &=   \angleb{(\Delta U)^2}_{NpT}
+ p^2\angleb{(\Delta V)^2}_{NpT} + 2p\angleb{\Delta U\Delta V}_{NpT} \\
&= \angleb{(\Delta U)^2}_{NVT} + \frac{k_BTK_T}{V}
\left( \frac{W}{K_T} - \frac{\angleb{\Delta U\Delta W}_{NVT}}{k_BTK_T}\right)^2
\nonumber \\
&+\frac{p^2Vk_BT}{K_T} + \frac{2p}{K_T}\left(-k_BTW 
+ \angleb{\Delta U\Delta W}_{NVT}\right) \\
&= \angleb{(\Delta U)^2}_{NVT} 
+ \frac{k_BT}{VK_T} \left(W^2-\frac{2W\angleb{\Delta U\Delta W}_{NVT}}{k_BT}
+\frac{\angleb{\Delta U\Delta W}_{NVT}^2}{(k_BT)^2} \right. \nonumber \\
&+ \left. (pV)^2 -2pVW + \frac{2pV\angleb{\Delta U\Delta W}_{NVT}}{k_BT}\right)\,.
\end{align}
Again using $pV=Nk_BT+W$ allows some simplication:

\begin{align}
\angleb{(\Delta H_{\rm ex})^2}_{NpT} &= \angleb{(\Delta U)^2}_{NVT} 
+ \frac{k_BT}{VK_T} \left(W^2-\frac{2W\angleb{\Delta U\Delta W}_{NVT}}{k_BT}
+\frac{\angleb{\Delta U\Delta W}_{NVT}^2}{(k_BT)^2} \right. \nonumber \\
& +  W^2 + 2WNk_BT + (Nk_BT)^2 - 2W^2 - 2Nk_BTW \nonumber \\
& + \left. \frac{2W\angleb{\Delta U\Delta W}_{NVT}}{k_BT} + 2N\angleb{\Delta U\Delta W}_{NVT}  \right) \\
&=  \angleb{(\Delta U)^2}_{NVT} + \frac{\angleb{\Delta U\Delta W}_{NVT}^2}{k_BTVK_T}
+\frac{N^2(k_BT)^3}{VK_T} + \frac{2Nk_BT}{VK_T}\angleb{\Delta U\Delta W}_{NVT}.
\end{align}
Now we can form the $H_{\rm ex}V$ correlation coefficient,

\begin{align}
R_{H_{\rm ex}V,NpT} &= \frac{\angleb{\Delta H_{\rm ex}\Delta V}_{NpT}}
{
\sqrt{\angleb{(\Delta H_{\rm ex})^2}_{NpT}}
\sqrt{\angleb{(\Delta V)^2}_{NpT}}
} \\
&= \frac{(\angleb{\Delta U\Delta W}_{NVT} + N(k_BT)^2)/K_T}
{
\sqrt{ \angleb{(\Delta U)^2}_{NVT} + 
\frac{\angleb{\Delta U\Delta W}_{NVT}^2}{k_BTVK_T}
+\frac{N^2(k_BT)^3}{VK_T} + \frac{2Nk_BT}{VK_T}\angleb{\Delta U\Delta W}_{NVT}}
\sqrt{\frac{Vk_BT}{K_T}}
} \\
&= \frac{\angleb{\Delta U\Delta W}_{NVT} + N(k_BT)^2}
{\sqrt{K_TVk_BT\angleb{(\Delta U)^2}_{NVT} + 
\angleb{\Delta U\Delta W}_{NVT}^2
+ N^2(k_BT)^4 + 2N\angleb{\Delta U\Delta W}_{NVT}(k_BT)^2}} \\
&= \frac{a}{\sqrt{b^2+a^2}} = \frac{1}{\sqrt{1+b^2/a^2}}
\end{align}
where $a=\angleb{\Delta U\Delta W}_{NVT} + N(k_BT)^2$ and $b^2=K_TVk_BT\angleb{(\Delta U)^2}_{NVT}$.


\begin{thebibliography}{99}


\expandafter\ifx\csname natexlab\endcsname\relax\def\natexlab#1{#1}\fi
\expandafter\ifx\csname bibnamefont\endcsname\relax
  \def\bibnamefont#1{#1}\fi
\expandafter\ifx\csname bibfnamefont\endcsname\relax
  \def\bibfnamefont#1{#1}\fi
\expandafter\ifx\csname citenamefont\endcsname\relax
  \def\citenamefont#1{#1}\fi
\expandafter\ifx\csname url\endcsname\relax
  \def\url#1{\texttt{#1}}\fi
\expandafter\ifx\csname urlprefix\endcsname\relax\def\urlprefix{URL }\fi
\providecommand{\bibinfo}[2]{#2}
\providecommand{\eprint}[2][]{\url{#2}}

\bibitem[{\citenamefont{Pedersen
  et~al.}(2008{\natexlab{a}})\citenamefont{Pedersen, Bailey, Schr{\o}der, and
  Dyre}}]{Pedersen/others:2008}
\bibinfo{author}{\bibfnamefont{U.~R.} \bibnamefont{Pedersen}},
  \bibinfo{author}{\bibfnamefont{N.~P.} \bibnamefont{Bailey}},
  \bibinfo{author}{\bibfnamefont{T.~B.} \bibnamefont{Schr{\o}der}},
  \bibnamefont{and} \bibinfo{author}{\bibfnamefont{J.~C.} \bibnamefont{Dyre}},
  \bibinfo{journal}{Phys. Rev. Lett.} \textbf{\bibinfo{volume}{100}},
  \bibinfo{pages}{015701} (\bibinfo{year}{2008}{\natexlab{a}}).

\bibitem[{\citenamefont{Pedersen
  et~al.}(2008{\natexlab{b}})\citenamefont{Pedersen, Christensen, Schr{\o}der,
  and Dyre}}]{Pedersen/others:2008a}
\bibinfo{author}{\bibfnamefont{U.~R.} \bibnamefont{Pedersen}},
  \bibinfo{author}{\bibfnamefont{T.}~\bibnamefont{Christensen}},
  \bibinfo{author}{\bibfnamefont{T.~B.} \bibnamefont{Schr{\o}der}},
  \bibnamefont{and} \bibinfo{author}{\bibfnamefont{J.~C.} \bibnamefont{Dyre}},
  \bibinfo{journal}{Phys. Rev. E} \textbf{\bibinfo{volume}{77}},
  \bibinfo{pages}{011201} (\bibinfo{year}{2008}{\natexlab{b}}).

\bibitem[{\citenamefont{Bailey et~al.}(2008{\natexlab{c}})\citenamefont{Bailey,
  Christensen, Jakobsen, Niss, Olsen, Pedersen, Schr{\o}der, and
  Dyre}}]{Bailey/others:2008a}
\bibinfo{author}{\bibfnamefont{N.~P.} \bibnamefont{Bailey}},
  \bibinfo{author}{\bibfnamefont{T.}~\bibnamefont{Christensen}},
  \bibinfo{author}{\bibfnamefont{B.}~\bibnamefont{Jakobsen}},
  \bibinfo{author}{\bibfnamefont{K.}~\bibnamefont{Niss}},
  \bibinfo{author}{\bibfnamefont{N.~B.} \bibnamefont{Olsen}},
  \bibinfo{author}{\bibfnamefont{U.~R.} \bibnamefont{Pedersen}},
  \bibinfo{author}{\bibfnamefont{T.~B.} \bibnamefont{Schr{\o}der}},
  \bibnamefont{and} \bibinfo{author}{\bibfnamefont{J.~C.} \bibnamefont{Dyre}},
  \bibinfo{journal}{J. Phys.: Condens. Matter} \textbf{\bibinfo{volume}{20}},
  \bibinfo{pages}{244113} (\bibinfo{year}{2008}{\natexlab{c}}).

\bibitem[{\citenamefont{Bailey et~al.}(2008{\natexlab{a}})\citenamefont{Bailey,
  Pedersen, Gnan, Schr{\o}der, and Dyre}}]{Bailey/others:2008b}
\bibinfo{author}{\bibfnamefont{N.~P.} \bibnamefont{Bailey}},
  \bibinfo{author}{\bibfnamefont{U.~R.} \bibnamefont{Pedersen}},
  \bibinfo{author}{\bibfnamefont{N.}~\bibnamefont{Gnan}},
  \bibinfo{author}{\bibfnamefont{T.~B.} \bibnamefont{Schr{\o}der}},
  \bibnamefont{and} \bibinfo{author}{\bibfnamefont{J.~C.} \bibnamefont{Dyre}},
  \bibinfo{journal}{J. Chem. Phys.} \textbf{\bibinfo{volume}{129}},
  \bibinfo{pages}{184507} (\bibinfo{year}{2008}{\natexlab{a}}),
  \bibinfo{note}{(Paper I)}.

\bibitem[{\citenamefont{Bailey et~al.}(2008{\natexlab{b}})\citenamefont{Bailey,
  Pedersen, Gnan, Schr{\o}der, and Dyre}}]{Bailey/others:2008c}
\bibinfo{author}{\bibfnamefont{N.~P.} \bibnamefont{Bailey}},
  \bibinfo{author}{\bibfnamefont{U.~R.} \bibnamefont{Pedersen}},
  \bibinfo{author}{\bibfnamefont{N.}~\bibnamefont{Gnan}},
  \bibinfo{author}{\bibfnamefont{T.~B.} \bibnamefont{Schr{\o}der}},
  \bibnamefont{and} \bibinfo{author}{\bibfnamefont{J.~C.} \bibnamefont{Dyre}},
  \bibinfo{journal}{J. Chem. Phys.} \textbf{\bibinfo{volume}{129}},
  \bibinfo{pages}{184508} (\bibinfo{year}{2008}{\natexlab{b}}),
  \bibinfo{note}{(Paper II)}.

\bibitem[{\citenamefont{Allen and Tildesley}(1987)}]{Allen/Tildesley:1987}
\bibinfo{author}{\bibfnamefont{M.~P.} \bibnamefont{Allen}} \bibnamefont{and}
  \bibinfo{author}{\bibfnamefont{D.~J.} \bibnamefont{Tildesley}},
  \emph{\bibinfo{title}{Computer Simulation of Liquids}}
  (\bibinfo{publisher}{Clarendon Press, Oxford}, \bibinfo{year}{1987}).

\bibitem[{\citenamefont{Dzugutov}(1992)}]{Dzugutov:1992}
\bibinfo{author}{\bibfnamefont{M.}~\bibnamefont{Dzugutov}},
  \bibinfo{journal}{Phys. Rev. A} \textbf{\bibinfo{volume}{46}},
  \bibinfo{pages}{R2984} (\bibinfo{year}{1992}).

\bibitem{IV} N. Gnan, T. B. Schr{\o}der, U. R. Pedersen, N. P. Bailey, and J. C. Dyre, arXiv:0905.3497 (2008) (Paper IV, J. Chem. Phys., to appear).

\bibitem[{\citenamefont{Chisolm and Wallace}(2001)}]{Chisolm/Wallace:2001}
\bibinfo{author}{\bibfnamefont{E.~D.} \bibnamefont{Chisolm}} \bibnamefont{and}
  \bibinfo{author}{\bibfnamefont{D.~C.} \bibnamefont{Wallace}},
  \bibinfo{journal}{J. Phys.: Condens. Matter} \textbf{\bibinfo{volume}{13}},
  \bibinfo{pages}{R739} (\bibinfo{year}{2001}).

\bibitem[{\citenamefont{Oster et~al.}(1971)\citenamefont{Oster, Perelson, and
  Katchalsky}}]{Oster/Perelson/Katchalsky:1971}
\bibinfo{author}{\bibfnamefont{G.} \bibnamefont{Oster}},
  \bibinfo{author}{\bibfnamefont{A.} \bibnamefont{Perelson}},
  \bibnamefont{and}
  \bibinfo{author}{\bibfnamefont{A.}~\bibnamefont{Katchalsky}},
  \bibinfo{journal}{Nature} \textbf{\bibinfo{volume}{234}},
  \bibinfo{pages}{393} (\bibinfo{year}{1971}).


\bibitem[{\citenamefont{Klein}(1919)}]{kle19}
\bibinfo{author}{\bibfnamefont{O.}~\bibnamefont{Klein}},
  \bibinfo{journal}{Medd. Vetenskapsakad. Nobelinst.}
  \textbf{\bibinfo{volume}{5}}, \bibinfo{pages}{No. 6} (\bibinfo{year}{1919}).

\bibitem[{\citenamefont{Berlin and Montroll}(1952)}]{ber52}
\bibinfo{author}{\bibfnamefont{T.~H.} \bibnamefont{Berlin}} \bibnamefont{and}
  \bibinfo{author}{\bibfnamefont{E.~W.} \bibnamefont{Montroll}},
  \bibinfo{journal}{J. Chem. Phys.} \textbf{\bibinfo{volume}{20}},
  \bibinfo{pages}{75} (\bibinfo{year}{1952}).

\bibitem{hoo70} W. G. Hoover, M. Ross, K. W. Johnson, D. Henderson, J. A. Barker, and B. C. Brown, J. Chem. Phys. {\bf 52}, 4931 (1970).

\bibitem{hoo71} W. G. Hoover, S. G. Gray, and K.W. Johnson, J. Chem. Phys. {\bf 55}, 1128 (1971).

\bibitem{hiw74} Y. Hiwatari, H. Matsuda, T. Ogawa, N. Ogita, and A. Ueda, Prog. Theor. Phys. {\bf 52}, 1105 (1974).

\bibitem{benamotz03} D. Ben-Amotz and G. J. Stell, J. Chem. Phys. {\bf 119}, 10777 (2003).

\bibitem{dem04} C. DeMichele, F. Sciortino, and A. Coniglio, J. Phys.: Condens. Matter {\bf 16}, L489 (2004).

\bibitem{ram09} P. E. Ramirez-Gonzalez and M. Medina-Noyola, J. Phys.: Condens. Matter {\bf 21}, 075101 (2009).

\bibitem{sti75} S. M. Stishov, Sov. Phys. Usp. {\bf 17}, 625 (1975).

\bibitem{wee83} J. D. Weeks and J. Q. Broughton, J. Chem. Phys. {\bf 78}, 4197 (1983).

\bibitem[{\citenamefont{Hansen and McDonald}(1986)}]{Hansen/McDonald:1986}
\bibinfo{author}{\bibfnamefont{J.~P.} \bibnamefont{Hansen}} \bibnamefont{and}
  \bibinfo{author}{\bibfnamefont{I.~R.} \bibnamefont{McDonald}},
  \emph{\bibinfo{title}{{Theory of Simple Liquids}}}
  (\bibinfo{publisher}{Academic Press, New York}, \bibinfo{year}{1986}),
  \bibinfo{edition}{2nd} ed.

\bibitem[{\citenamefont{Schr{\o}der
  et~al.}(2008{\natexlab{b}})\citenamefont{Schr{\o}der, Pedersen, and
  Dyre}}]{Schroder/Pedersen/Dyre:2008a}
\bibinfo{author}{\bibfnamefont{T.~B.} \bibnamefont{Schr{\o}der}},
  \bibinfo{author}{\bibfnamefont{U.~R.} \bibnamefont{Pedersen}},
  \bibnamefont{and} \bibinfo{author}{\bibfnamefont{J.~C.} \bibnamefont{Dyre}}
  (\bibinfo{year}{2008}{\natexlab{b}}), \eprint{arXiv:0803.2199}.

\bibitem[{\citenamefont{Schr{\o}der
  et~al.}(2008{\natexlab{a}})\citenamefont{Schr{\o}der, Pedersen, Bailey,
  Toxv{\ae}rd, and Dyre}}]{Schroder/others:2008}
\bibinfo{author}{\bibfnamefont{T.~B.} \bibnamefont{Schr{\o}der}},
  \bibinfo{author}{\bibfnamefont{U.~R.} \bibnamefont{Pedersen}},
  \bibinfo{author}{\bibfnamefont{N.~P.} \bibnamefont{Bailey}},
  \bibinfo{author}{\bibfnamefont{S.}~\bibnamefont{Toxv{\ae}rd}},
  \bibnamefont{and} \bibinfo{author}{\bibfnamefont{J.~C.} \bibnamefont{Dyre}}
  (\bibinfo{year}{2008}{\natexlab{a}}), \eprint{arXiv:0812.4960 [Phys. Rev. E, accepted]}.

\bibitem[{\citenamefont{Alba-Simionesco
  et~al.}(2004)\citenamefont{Alba-Simionesco, Cailliaux, Alegria, and
  Tarjus}}]{Alba-Simionesco/others:2004}
\bibinfo{author}{\bibfnamefont{C.}~\bibnamefont{Alba-Simionesco}},
  \bibinfo{author}{\bibfnamefont{A.}~\bibnamefont{Cailliaux}},
  \bibinfo{author}{\bibfnamefont{A.}~\bibnamefont{Alegria}}, \bibnamefont{and}
  \bibinfo{author}{\bibfnamefont{G.}~\bibnamefont{Tarjus}},
  \bibinfo{journal}{Europhys. Lett.} \textbf{\bibinfo{volume}{68}},
  \bibinfo{pages}{58} (\bibinfo{year}{2004}).

\bibitem[{\citenamefont{Roland et~al.}(2005)\citenamefont{Roland,
  Hensel-Bielowka, Paluch, and Casalini}}]{Roland/others:2005}
\bibinfo{author}{\bibfnamefont{C.~M.} \bibnamefont{Roland}},
  \bibinfo{author}{\bibfnamefont{S.}~\bibnamefont{Hensel-Bielowka}},
  \bibinfo{author}{\bibfnamefont{M.}~\bibnamefont{Paluch}}, \bibnamefont{and}
  \bibinfo{author}{\bibfnamefont{R.}~\bibnamefont{Casalini}},
  \bibinfo{journal}{Rep. Prog. Phys.} \textbf{\bibinfo{volume}{68}},
  \bibinfo{pages}{1405} (\bibinfo{year}{2005}).

\bibitem[{\citenamefont{Grzybowski et~al.}(2009)\citenamefont{Grzybowski,
  Paluch, and Grzybowska}}]{Grzybowski/Paluch/Grzybowska:2009}
\bibinfo{author}{\bibfnamefont{A.}~\bibnamefont{Grzybowski}},
  \bibinfo{author}{\bibfnamefont{M.}~\bibnamefont{Paluch}}, \bibnamefont{and}
  \bibinfo{author}{\bibfnamefont{K.}~\bibnamefont{Grzybowska}},
  \bibinfo{journal}{J. Phys. Chem. B} \textbf{\bibinfo{volume}{113}},
  \bibinfo{pages}{7419} (\bibinfo{year}{2009}).

\bibitem{mol09} V. Molinero and E. B. Moore, J. Phys. Chem. B. {\bf 113}, 4008 (2009).

\bibitem{jagla} E. A. Jagla, J. Chem. Phys. {\bf 111}, 8980 (1999).

\bibitem[{\citenamefont{Lewis and Wahnstr{\"o}m}(1994)}]{Lewis/Wahnstrom:1994}
\bibinfo{author}{\bibfnamefont{L.~J.} \bibnamefont{Lewis}} \bibnamefont{and}
  \bibinfo{author}{\bibfnamefont{G.}~\bibnamefont{Wahnstr{\"o}m}},
  \bibinfo{journal}{Phys. Rev. E} \textbf{\bibinfo{volume}{50}},
  \bibinfo{pages}{3865} (\bibinfo{year}{1994}).

\bibitem[{\citenamefont{Berendsen et~al.}(1987)\citenamefont{Berendsen,
  Grigera, and Straatsma}}]{Berendsen/Grigera/Straatsma:1987}
\bibinfo{author}{\bibfnamefont{H.~J.~C.} \bibnamefont{Berendsen}},
  \bibinfo{author}{\bibfnamefont{J.~R.} \bibnamefont{Grigera}},
  \bibnamefont{and} \bibinfo{author}{\bibfnamefont{T.~P.}
  \bibnamefont{Straatsma}}, \bibinfo{journal}{J. Phys. Chem.}
  \textbf{\bibinfo{volume}{91}}, \bibinfo{pages}{6269} (\bibinfo{year}{1987}).

\bibitem[{\citenamefont{Goldstein}(1969)}]{Goldstein:1969}
\bibinfo{author}{\bibfnamefont{M.}~\bibnamefont{Goldstein}},
  \bibinfo{journal}{J. Chem. Phys.} \textbf{\bibinfo{volume}{51}},
  \bibinfo{pages}{3728} (\bibinfo{year}{1969}).

\bibitem[{\citenamefont{Stillinger and Weber}(1983)}]{Stillinger/Weber:1983}
\bibinfo{author}{\bibfnamefont{F.~H.} \bibnamefont{Stillinger}}
  \bibnamefont{and} \bibinfo{author}{\bibfnamefont{T.~A.} \bibnamefont{Weber}},
  \bibinfo{journal}{Phys. Rev. A} \textbf{\bibinfo{volume}{28}},
  \bibinfo{pages}{2408} (\bibinfo{year}{1983}).

\bibitem[{\citenamefont{Stillinger}(1995)}]{Stillinger:1995}
\bibinfo{author}{\bibfnamefont{F.~H.} \bibnamefont{Stillinger}},
  \bibinfo{journal}{Science} \textbf{\bibinfo{volume}{267}},
  \bibinfo{pages}{1935} (\bibinfo{year}{1995}).

\bibitem[{\citenamefont{Schr{\o}der et~al.}(2000)\citenamefont{Schr{\o}der,
  Sastry, Dyre, and Glotzer}}]{Schroder/others:2000}
\bibinfo{author}{\bibfnamefont{T.~B.} \bibnamefont{Schr{\o}der}},
  \bibinfo{author}{\bibfnamefont{S.}~\bibnamefont{Sastry}},
  \bibinfo{author}{\bibfnamefont{J.~C.} \bibnamefont{Dyre}}, \bibnamefont{and}
  \bibinfo{author}{\bibfnamefont{S.~C.} \bibnamefont{Glotzer}},
  \bibinfo{journal}{J. Chem. Phys.} \textbf{\bibinfo{volume}{112}},
  \bibinfo{pages}{9834} (\bibinfo{year}{2000}).

\bibitem[{\citenamefont{Lebowitz et~al.}(1967)\citenamefont{Lebowitz, Percus,
  and Verlet}}]{Lebowitz/Perkus/Verlet:1967}
\bibinfo{author}{\bibfnamefont{J.~L.} \bibnamefont{Lebowitz}},
  \bibinfo{author}{\bibfnamefont{J.~K.} \bibnamefont{Perkus}},
  \bibnamefont{and} \bibinfo{author}{\bibfnamefont{L.}~\bibnamefont{Verlet}},
  \bibinfo{journal}{Phys. Rev.} \textbf{\bibinfo{volume}{153}},
  \bibinfo{pages}{250} (\bibinfo{year}{1967}).

\bibitem[{\citenamefont{Born and Huang}(1954)}]{Born/Huang:1954}
\bibinfo{author}{\bibfnamefont{M.}~\bibnamefont{Born}} \bibnamefont{and}
  \bibinfo{author}{\bibfnamefont{K.}~\bibnamefont{Huang}},
  \emph{\bibinfo{title}{{Dynamical Theory of Crystal Lattices}}}
  (\bibinfo{publisher}{Oxford University Press}, \bibinfo{year}{1954}).

\bibitem[{\citenamefont{Ashcroft and Mermin}(1976)}]{Ashcroft/Mermin:1976}
\bibinfo{author}{\bibfnamefont{N.~W.} \bibnamefont{Ashcroft}} \bibnamefont{and}
  \bibinfo{author}{\bibfnamefont{N.~D.} \bibnamefont{Mermin}},
  \emph{\bibinfo{title}{{Solid State Physics}}} (\bibinfo{publisher}{Holt, Rinehart and Wiston, New York
  College}, \bibinfo{year}{1976}).

\bibitem[{\citenamefont{Wallace}(1972)}]{Wallace:1972}
\bibinfo{author}{\bibfnamefont{D.~C.} \bibnamefont{Wallace}},
  \emph{\bibinfo{title}{{Thermodynamics of Crystals}}}
  (\bibinfo{publisher}{Dover}, \bibinfo{year}{1972}).

\bibitem[{\citenamefont{Casalini et~al.}(2006)\citenamefont{Casalini, Mohanty,
  and Roland}}]{Casalini/Mohanty/Roland:2006}
\bibinfo{author}{\bibfnamefont{R.}~\bibnamefont{Casalini}},
  \bibinfo{author}{\bibfnamefont{U.}~\bibnamefont{Mohanty}}, \bibnamefont{and}
  \bibinfo{author}{\bibfnamefont{C.~M.} \bibnamefont{Roland}},
  \bibinfo{journal}{J. Chem. Phys.} \textbf{\bibinfo{volume}{125}},
  \bibinfo{pages}{014505} (\bibinfo{year}{2006}).

\bibitem[{\citenamefont{Feldman and Casalini}(2006)}]{Roland/Feldman/Casalini:2006}
\bibinfo{author}{\bibfnamefont{C.~M. Roland,~ J.~L.} \bibnamefont{Feldman,}}
  \bibnamefont{and} \bibinfo{author}{\bibfnamefont{R.}~\bibnamefont{Casalini}},
  \bibinfo{journal}{J. Non-Cryst. Solids} \textbf{\bibinfo{volume}{352}},
  \bibinfo{pages}{4895} (\bibinfo{year}{2006}).

\bibitem[{\citenamefont{Roland and Casalini}(2007)}]{Roland/Casalini:2007}
\bibinfo{author}{\bibfnamefont{C.~M.} \bibnamefont{Roland}} \bibnamefont{and}
  \bibinfo{author}{\bibfnamefont{R.}~\bibnamefont{Casalini}},
  \bibinfo{journal}{J. Phys.: Condens. Matter} \textbf{\bibinfo{volume}{19}},
  \bibinfo{pages}{205118} (\bibinfo{year}{2007}).

\bibitem[{\citenamefont{Paynter}(1961)}]{Paynter:1961}
\bibinfo{author}{\bibfnamefont{H.}~\bibnamefont{Paynter}},
  \emph{\bibinfo{title}{Analysis and Design of Engineering Systems}}
  (\bibinfo{publisher}{MIT, Cambridge, Mass.}, \bibinfo{year}{1961}).

\bibitem[{\citenamefont{Oster et~al.}(1973)\citenamefont{Oster, Perelson, and
  Katchalsky}}]{Oster/Perelson/Katchalsky:1973}
\bibinfo{author}{\bibfnamefont{G.~F.} \bibnamefont{Oster}},
  \bibinfo{author}{\bibfnamefont{A.~S.} \bibnamefont{Perelson}},
  \bibnamefont{and}
  \bibinfo{author}{\bibfnamefont{A.}~\bibnamefont{Katchalsky}},
  \bibinfo{journal}{Quarterly Rev. Biophys.} \textbf{\bibinfo{volume}{6}},
  \bibinfo{pages}{1} (\bibinfo{year}{1973}).

\bibitem[{\citenamefont{Christiansen}(1978)}]{Christiansen:1978}
\bibinfo{author}{\bibfnamefont{P.~V.} \bibnamefont{Christiansen}},
  \emph{\bibinfo{title}{{Dynamik og diagrammer}}} (\bibinfo{year}{1978}),
  \bibinfo{note}{{IMFUFA text No. 8, Roskilde}}.

\bibitem[{\citenamefont{Christiansen}(1979)}]{Christiansen:1979}
\bibinfo{author}{\bibfnamefont{P.~V.} \bibnamefont{Christiansen}},
  \emph{\bibinfo{title}{{Semiotik og systemegenskaber}}}
  (\bibinfo{year}{1979}), \bibinfo{note}{{IMFUFA text No. 22, Roskilde}}.

\bibitem[{\citenamefont{Mikulecky}(1993)}]{Mikulecky:1993}
\bibinfo{author}{\bibfnamefont{D.~C.} \bibnamefont{Mikulecky}},
  \emph{\bibinfo{title}{Applications of network thermodynamics to problems in
  biomedical engineering}} (\bibinfo{publisher}{New York University, New York},
  \bibinfo{year}{1993}).

\bibitem[{\citenamefont{Karnopp et~al.}(2006)\citenamefont{Karnopp, Margolis,
  and Rosenberg}}]{Karnopp/Margolis/Rosenberg:2006}
\bibinfo{author}{\bibfnamefont{D.~C.} \bibnamefont{Karnopp}},
  \bibinfo{author}{\bibfnamefont{D.~L.} \bibnamefont{Margolis}},
  \bibnamefont{and} \bibinfo{author}{\bibfnamefont{R.~C.}
  \bibnamefont{Rosenberg}}, \emph{\bibinfo{title}{System Dynamics: Modeling and
  Simulation of Mechatronic Systems}} (\bibinfo{publisher}{Wiley, New York},
  \bibinfo{year}{2006}).

\bibitem[{\citenamefont{Ellegaard et~al.}(2007)\citenamefont{Ellegaard,
  Christensen, Christiansen, Olsen, Pedersen, Schr{\o}der, and
  Dyre}}]{Ellegaard/others:2007}
\bibinfo{author}{\bibfnamefont{N.~L.} \bibnamefont{Ellegaard}},
  \bibinfo{author}{\bibfnamefont{T.}~\bibnamefont{Christensen}},
  \bibinfo{author}{\bibfnamefont{P.~V.} \bibnamefont{Christiansen}},
  \bibinfo{author}{\bibfnamefont{N.~B.} \bibnamefont{Olsen}},
  \bibinfo{author}{\bibfnamefont{U.~R.} \bibnamefont{Pedersen}},
  \bibinfo{author}{\bibfnamefont{T.~B.} \bibnamefont{Schr{\o}der}},
  \bibnamefont{and} \bibinfo{author}{\bibfnamefont{J.~C.} \bibnamefont{Dyre}},
  \bibinfo{journal}{J. Chem. Phys.} \textbf{\bibinfo{volume}{126}},
  \bibinfo{pages}{074502} (\bibinfo{year}{2007}).

\bibitem[{\citenamefont{Christensen and Dyre}(2008)}]{Christensen/Dyre:2008}
\bibinfo{author}{\bibfnamefont{T.}~\bibnamefont{Christensen}} \bibnamefont{and}
  \bibinfo{author}{\bibfnamefont{J.~C.} \bibnamefont{Dyre}},
  \bibinfo{journal}{Phys. Rev. E} \textbf{\bibinfo{volume}{78}},
  \bibinfo{pages}{021501} (\bibinfo{year}{2008}).

\bibitem[{\citenamefont{Berendsen et~al.}(1995)\citenamefont{Berendsen, van~der
  Spoel, and van Drunen}}]{Berendsen/vanderSpoel/vanDrunen:1995}
\bibinfo{author}{\bibfnamefont{H.~J.~C.} \bibnamefont{Berendsen}},
  \bibinfo{author}{\bibfnamefont{D.}~\bibnamefont{van~der Spoel}},
  \bibnamefont{and} \bibinfo{author}{\bibfnamefont{R.}~\bibnamefont{van
  Drunen}}, \bibinfo{journal}{Comp. Phys. Comm.} \textbf{\bibinfo{volume}{91}},
  \bibinfo{pages}{43} (\bibinfo{year}{1995}).

\bibitem[{\citenamefont{Lindahl et~al.}(2001)\citenamefont{Lindahl, Hess, and
  van~der Spoel}}]{Lindahl/Hess/vanderSpoel:2001}
\bibinfo{author}{\bibfnamefont{E.}~\bibnamefont{Lindahl}},
  \bibinfo{author}{\bibfnamefont{B.}~\bibnamefont{Hess}}, \bibnamefont{and}
  \bibinfo{author}{\bibfnamefont{D.}~\bibnamefont{van~der Spoel}},
  \bibinfo{journal}{J. Mol. Mod.} \textbf{\bibinfo{volume}{7}},
  \bibinfo{pages}{306} (\bibinfo{year}{2001}).

\bibitem[{\citenamefont{Nos\'e}(1984)}]{Nose:1984}
\bibinfo{author}{\bibfnamefont{S.} \bibnamefont{Nos\'e}},
  \bibinfo{journal}{Mol. Phys.} \textbf{\bibinfo{volume}{52}},
  \bibinfo{pages}{255} (\bibinfo{year}{1984}).

\bibitem[{\citenamefont{Hoover}(1985)}]{Hoover:1985}
\bibinfo{author}{\bibfnamefont{W.~G.} \bibnamefont{Hoover}},
  \bibinfo{journal}{Phys. Rev. A} \textbf{\bibinfo{volume}{31}},
  \bibinfo{pages}{1695} (\bibinfo{year}{1985}).

\bibitem[{\citenamefont{Hess et~al.}(1997)\citenamefont{Hess, Bekker,
  Berendsen, and Fraaije}}]{Hess/others:1997}
\bibinfo{author}{\bibfnamefont{B.}~\bibnamefont{Hess}},
  \bibinfo{author}{\bibfnamefont{H.}~\bibnamefont{Bekker}},
  \bibinfo{author}{\bibfnamefont{H.~J.~C.} \bibnamefont{Berendsen}},
  \bibnamefont{and} \bibinfo{author}{\bibfnamefont{J.~G. E.~M.}
  \bibnamefont{Fraaije}}, \bibinfo{journal}{J. Comp. Chem.}
  \textbf{\bibinfo{volume}{18}}, \bibinfo{pages}{1463} (\bibinfo{year}{1997}).

\bibitem{cos09} D. Coslovich and C. M. Roland, J. Chem. Phys. {\bf 130}, 014508 (2009).


\end{thebibliography}
\end{document}